# Toward the use of temporary tattoo electrodes for impedancemetric respiration monitoring and other electrophysiological recordings

S. Taccola*[1,2], A. Poliziani[1], D. Santonocito[3], A. Mondini[1], C. Denk[3], A. N. Ide[3], M. Oberparleiter[3], F. Greco*[1,4], V. Mattoli*[1]

[1] Center for Micro-BioRobotics, Istituto Italiano di Tecnologia, Viale Rinaldo Piaggio 34, 56025 Pontedera, Pisa, Italy.

[2] Future Manufacturing Processes Research Group, School of Mechanical Engineering, Faculty of Engineering, University of Leeds, Leeds LS2 9JT, United Kingdom.

[3] Emerging Application Department, MED-EL Elektromedizinische Geräte Gesellschaft m.b.H., Fürstenweg 77a, 6020 Innsbruck, Austria

[4] Institute of Solid State Physics, NAWI Graz, Graz University of Technology, Petersgasse 16, 8010 Graz, Austria.

*E-mail: S.Taccola@leeds.ac.uk; francesco.greco@tugraz.at; virgilio.mattoli@iit.it

ABSTRACT: Development of dry, ultra-conformable and unperceivable temporary tattoo electrodes (TTEs), based on the ink-jet printing of PEDOT:PSS on top of commercially available temporary tattoo paper, has gained increasing attention as a new and promising technology for electrophysiological recordings on skin. In this work we present a TTEs epidermal sensor for real time monitoring of respiration through transthoracic impedance measurements, exploiting a new design based on the application of soft screen printed Ag ink and magnetic interlink, that guarantees a repositionable, long-term stable and robust interconnection of TTEs with external "docking" devices. The efficiency of the TTE and the proposed interconnection strategy under stretching (up to 10%) and over time (up to 96 hours) has been verified on a dedicated experimental setup and on humans, fulfilling the proposed specific application of transthoracic impedance measurements. The proposed approach makes this technology suitable for large-scale production and suitable not only for the specific use case presented, but also for real time monitoring of different bio-electric signals, as demonstrated through specific proof of concept demonstrators.

# 1. Introduction

The terms "epidermal electronics", "skin-like", and "electronic tattoos" refer to a class of electronic devices with physical properties (i.e. elastic modulus, thickness, bending stiffness and areal mass density) that approximate those of the epidermis, thereby enabling a non-invasive but intimate coupling with the complex features of the skin [1, 2]. Electronic devices conformally mounted directly on the skin can provide a versatile means to acquire information about the body through the monitoring of biologically relevant chemical and physical variables, which could be tracked for continuous health monitoring, as well as for robotic feedback and control, prosthetics and rehabilitation, and human/computer interfaces [3-5].

Superior adhesion of devices to the skin is achieved by either making substrates soft and thin. Within this framework, one of the most promising approaches for conformable electronics is the development of multilayer organic thin-film structures, assembled on ultrathin polymeric membranes [3, 6-8]. These structures are subsequently laminated onto the epidermis, conformally adhering to the skin thanks to physical interactions alone, i.e. van der Waals forces. Such conformability maximizes the contact area between the sensors and the skin, minimizing the contact impedance and facilitating signal transfer across the sensor-skin interface. Moreover, the reduction of the mismatch in the mechanical properties between skin and sensor, enables the sensor to follow skin displacement and deformation, minimizing interfacial slippage. As a result, motion artefacts affecting the performances of thick and rigid conventional sensors are mitigated in ultrathin ones, providing accurate, non-invasive, long-term, and continuous measurements [9-11]

Recently, commercially available temporary tattoo paper has attracted the attention of many researchers as a platform for device integration [12-21]. Epidermal electronics fabricated on top of decal transfer paper enables an easy, reliable, fast manipulation, release/transfer on the skin (or other target surfaces), which are crucial from an end-user standpoint; as in the case of standard temporary transfer tattoos, the release occurs by the dissolution of a water-soluble sacrificial layer, wetting the supporting paper sheet with water and pressing against the target surface (usually, the skin) for few seconds. In particular, temporary tattoo electrodes (TTEs) technology has been successfully demonstrated to work as a new and promising technology for surface electrophysiological recordings [14, 18, 19]. Within this framework disposable Ag/AgCl electrodes, which operate with an electrolytic gel coupled to skin to establish the electrical contact and reduce the skin impedance, are conventionally used either for clinical practice or research. Despite their intensive use, standard gelled electrodes show many drawbacks, such as discomfort, rigidity against skin compliance and limited signal stability mainly due to the drying of the gel over time[22]. Dry electrodes that can operate without gel and adhesive have been studied for many decades, with the main limitation of poor electrode-to-skin contact leading to higher impedance and more susceptibility to motion artefact. These issues may be addressed by using tattoo conductive polymer nanosheets recently proposed by Greco et al. [12, 14]; these ultrathin nanosheets, composed of conducting polymer complex poly(3,4-ethylenedioxythiophene) polystyrene sulfonate (PEDOT: PSS) printed on top of a thin (~500 nm) ethylcellulose layer released from a decal transfer paper, can provide ultra-conformability on a complex surface as skin, addressing the issue with lack of conformability and poor adhesion which occurs with standard dry electrodes. Very recently Ferrari *et al* proposed PEDOT:PSS ink-jet printed TTEs for the measurement of EMG on limbs and face, as well as ECG [14] and EEG recordings in a clinical environment, representing the first example ever of dry electrodes with full magnetoencephalography compatibility [23].

The interfacing of stretchable and skin-conformal devices with hard electronic components, such as interconnects, memory, energy devices and wireless electronics for data collection and communication, is of major interest and has attracted intensive worldwide research efforts [24, 25]. Several materials and interface design strategies, such as buckled [26, 27], serpentine-shaped [28], and micro-patterned [29] design, have been reported in literature to improve soft/hard material interfaces, preventing failure of electromechanical properties under mechanical deformation and achieving reliable device performance. A comprehensive review that covers the state-of-the art development in this field with a strong emphasis on multiscale soft–hard interface design has been recently published by Gong et al [25]. However, for what concerns the dry-electrode tattoo-like technology, the integration of easy-to-handle suitable interconnections with external devices remains somehow an open point that needs to be addressed, toward the application of that technology in long term (e.g. longer than few hours) real-time electrophysiological recordings applications. The present work we demonstrate the use of TTEs in a specific relevant real-word application, the respiration monitoring trough transthoracic bio impedance recording, proposing a new approach to suitable TTEs connections with external electronics devices, based on synergic application of soft screen printed Ag ink and magnetic interlink. that This design guarantees a stable, easy to handle interconnection mechanism that has been further tested with additional body signal recording proof-of-concept demonstrators.

## 2. Case study, ultrathin films and interconnection technology

The respiratory rate ($f_R$) is a very informative vital sign responding to a variety of stressors. In clinical environment, $f_R$ is a typically monitored health parameter, especially in case of chronic respiratory diseases, and also a predictor of potentially serious adverse events (e.g. cardiac arrest, sleep apnea, and infant death syndrome) [30]. Moreover, the monitoring of $f_R$ in occupational settings or sporting activities, can provide useful information to improve health, well-being, and safety. Indeed, $f_R$ is sensitive to physical effort, cognitive load, emotional stress, environmental challenges, pain, and discomfort, etc [31]. Continuous respiration monitoring can be achieved through different methods [31, 32] which include, for example, the use of airflow sensors for measuring the volume and/or the velocity of the inhaled and exhaled air during breathing [33], acoustic sensors measuring the sound generated by the air flowing through the patient's throat and airways [34], humidity sensors measuring the difference of water vapour contents between inhaled and exhaled air [35], and techniques based on chest wall movement analysis such as strain sensors and bio-impedance sensors [36, 37]. Within this framework, the revolution of stretchable electronics and sensors has recently brought significant advancements toward continuous monitoring of respiration signals in the era of real time and portable healthcare systems; a comprehensive review article providing an overview on the recent development of wearable stretchable physical sensors for this purpose has been recently published by Dinh et al [38].

Among the different measurement strategies, bio-impedance is a non-invasive and comfortable technique for monitoring respiratory parameters when measured in the thorax. The monitoring of $f_R$ using bio-impedance sensors is based on the cyclic impedance changes caused by the respiratory-induced chest wall movements and respiratory volume: during inspiration, the increase in the gas volume causes an increment of the electrical impedance of the lungs, while the expansion of the chest increases the length of the conductance paths [39]. It has been demonstrated that the relationship between this impedance change and the respiratory volume is approximately linear, providing a way to measure respiratory activity by sensing these changes [31, 39]. Bio-impedance sensors require skin-contact electrodes for signal acquisition from the body, positioned at the level of the upper thorax. Due to the aforementioned limitations of traditional Ag/AgCl electrodes typically used for bio-impedance, the continuous and long term (i.e. over more consecutive days or even few hours) monitoring of respiratory rate by mean of wearable devices, is still an open and challenging field where optimal solutions are missing. For these reasons, we proposed wearable tattoo sensor for real time respiration monitoring through transthoracic impedance measurement, as specific use case of clinical relevance for validation of the proposed approach. To doing this, a suitable and stable interconnection between TTEs and external electronic devices (see. Figure 1.a) is required.

Stable and conformable adhesion of inkjet-printed PEDOT:PSS TTEs onto skin on several locations on the body has been already demonstrated in our previous work, also at the microscopic level [14]. In this work we use a similar approach, but with different materials and assembly. In particular, we used another commercial decal transfer paper as the substrate for TTEs fabrication, to improve mechanical performances. A detailed description of the paper composition is reported in Materials and Methods section (see also Supporting Materials SI.1). The thickness of the supporting tattoo layer, which is released on the skin after the dissolution of the water-soluble sacrificial layer and the removal of the supporting paper, was measured to be ~1.7 μm. In order to establish the external electrical connection to the TTE sensor, a 25 μm thick polyimide sheet has been employed.

Therefore, a mechanical mismatch between materials of different thickness arises and causes the interface between the two materials to be very fragile. Moreover, any conductive tracks either deposited or printed across the mentioned interface will result in a breakage whenever the interface undergoes a certain mechanical stress (e.g. bending or stretching).

The failure at the interface between an ultrathin layer (conformally attached to the skin) and a thicker stiffer layer connected through thin conductive tracks can be caused by two different factors: flexural rigidity mismatch and the elastic modulus (Young's modulus) mismatch. The Flexural Rigidity D is defined as the bending moment (force couple) required to bend a structure per unit length per unit of curvature. It can be defined as the resistance offered by a plate structure while undergoing bending [40]:

$$\frac{Eh^3}{12(1-v^2)} = D$$

$$D\frac{d^2w}{dx^2} = -M$$

$$D = EI ,$$

where $E$ is the Young's modulus of the material, $h$ is the thickness of the beam, $v$ is the Poisson's ratio, $M$ is the internal bending moment of the beam, $d^2w/dx^2$ is the local curvature and $I$ is the area moment of inertia (also called second area moment) of the beam cross-section.

If we consider (as a typical example) a structure composed of two different layers characterized by different flexural rigidities and conformally attached to a curved surface, then a high concentration of the stress is generated at the interface between the two layers. This happens because of the different forces generated by the two parts in response to the same curvature. If the stress overcomes the maximum stress (in the thinner layer), breakages occur. Moreover, if we consider a structure composed of two different layers characterized by different Young's moduli, if strain is applied, then high stress is generated at the interface between the two layers. Consequently, cracks and breakages occur. Analogously, in more complex systems (e.g. multi-layers), high stress is generated at the different interfaces.

In order to address this issue, we followed a "graded material transition" approach, using an intermediate interconnection layer made up of a soft and stretchable silver conductor paste (commercial formulation of an elastomer and Ag nanoparticles, in the following referred as soft Ag ink or simply Ag Ink). The approach is schematically depicted in Figure 1.b. In details, interconnecting lines and pads, which join the PEDOT:PSS ultrathin electrodes with the thick external connectors, were deposited by screen printing of the soft Ag ink, both on the tattoo paper substrate (containing the PEDOT:PSS electrode) and on the polyimide film (for implementing external contacts/connections). Afterwards, the two parts (tattoo and polyimide part) have been assembled placing a drop of the same ink between the tracks on the two parts (see Figure 1.b for connection schematic section). The adoption of the soft Ag ink provided a compliant interface between layers having different thicknesses and mechanical properties, thus reducing the mismatch between the different layers both in terms of flexural rigidity and Young's moduli.

Finally, the connection of the tattoo sensor with external devices is achieved through a magnetically-enabled contact, in analogy to what presented by Rogers's group in [41]. In our approach, small magnets glued on the backside (polyimide side) of the Ag contact pads, squeeze the latters against conductive magnets on the external device, thus ensuring robust mechanical docking and reliable electrical contacts (see Figure 1.b). In order to keep the thicker part of the tattoo (polyimide and Ag ink tracks) on the skin, an additional thin bi-adhesive acrylic glue layer (the same provided in the tattoo paper kit) is used.

The overall configuration enables to exploit the conformal adhesion (mediated by Van der Walls force) of the PEDOT:PSS electrodes with the skin, while retaining a robust interface with external devices. The effectiveness of this solution has been tested for a specific case-study with clinical relevance as reported in the follow.

## 3. Results and discussion

### 3.1 TTEs for transthoracic impedance measurements design

Transthoracic impedance measurements are performed by using of a variable number of electrodes (either two or four) placed on the chest of the subject. In a typical four-electrodes (tetrapolar) configuration a high-frequency (typically 50 kHz) and low-amplitude current (less than 1 mA) is injected by two electrodes on the thorax, whereas the other two electrodes are used to record the impedance changes by measuring the voltage changes between them [39] (see Figure 1.c). Compared to the two-terminal measurement configuration, the four-electrode configuration yields a more accurate measurement because the sites of current injection and voltage measurement are physically separated [42]. In order to use TTEs for transthoracic impedance measurements, the design and the geometry of PEDOT:PSS electrodes and interconnections have been optimised for the proposed application, as illustrated in Figure 1.d. Four rectangular skin-contact electrodes were realized by inkjet printing on tattoo paper of a mixture of PEDOT:PSS aqueous dispersion and glycerol. Glycerol was added as a biocompatible additive improving conductivity and print quality of printed films [43]. PEDOT:PSS electrodes are electrically interconnected with a more thick and resistant polyamide part (for external connection), through flexible/stretchable tracks, deposited by screen printing of the stretchable Ag conductive paste. In order to increase the stretchability of Ag tracks along the longitudinal direction, a serpentine geometry was used. The use of a serpentine geometry is a good strategy to increase the stretchability of a conductive track in wearable and epidermal electronics [1]. The design of the polyimide central part has been optimized in order to guarantee the connection through magnetic docking with external connectors/devices. Four small (0.5 mm thick, 2.5 mm diameter) neodymium magnets (one for each pad) were fixed with glue on a plastic support to obtain a magnetic disk (Figure 1.d). The magnetic disk was finally attached to the on-tattoo mounted polyimide contact pad, on the opposite side of the Ag pads. Further details on fabrication are reported in Materials and Methods section (see also Supporting Materials SI.2). Magnets on tattoo are placed in a position that maximize the squeezing force for planar contacts between the contact pads and the pads on the external electrical connectors. Figure 1.e shows a tattoo released on the chest of a subject, and a magnetic connector used for signal acquisition. The magnetic interlocking system is able to sustain multiple connections/disconnections to the docking device without failure.

## 3.2 In lab stretching Test

TTEs have been designed to measure transthoracic impedance once placed on the thorax of the subject at the sternal angle, following the respiratory-induced rhythmic movement of the chest wall. The sternal angle is an ideal spot for bio-impedance measurements, as it is characterized by a low percentage of fat in the underlying skin layer, a thinner pectoralis muscle thickness, and a more stable position (respect to other points) also during high body movement situations since the bone is not part of any joint subject to roll, pitch and yaw.

Ideally the TTE should satisfy a series of requirements to be used in the proposed application, including, among other, the possibility to be stretched over the length of the chest expansion/contraction during the respiration, without losing the capability to record the impedance signals for the targeted amount of time, i.e. at least 24h for our application. Additionally, the TTE, thanks to the intrinsic conformability, should prevent (or limit) the artefacts originated by the fast change of the impedance at skin-electrode interface generated by body movements. The latters can indeed induce relative movements (sliding and local detachment) between electrode surface and skin surface, especially in case of bulky or poorly adherent electrodes

Quantitatively evaluation of these aspects could not directly be performed on human subjects, because it was considered not reliable due to the impossibility to have high repeatability of the movements, either between different subjects or on the same subject. Therefore, a customized test platform has been implemented to simulate the expansion/contraction that occur at the human chest (where the TTE prototypes are released), because of the inhalation/exhalation respectively. The testing equipment, whose setup is described in detail in the experimental section, is shown in Figure 2.a. Since specific and standard tests do not exist for this application, a customized experimental protocol for the evaluation of mechanical and electrical properties of the TTEs with time and deformation has been defined. The experimental protocol included three fundamental points: (i) the TTE shall be released onto a stretchable and conductive substrate that simulates the surface of the chest wall. (ii) The substrate shall be deformed in a way that simulates the deformation of the chest during respiration. (iii) Together with the above mentioned deformation, also a vertical acceleration shall also be considered to simulate real-life scenarios such as breathing during walking and/or running.

Consequently, conductive and stretchable rubber sheets have been selected as substrates onto which TTEs have been transferred as shown in Figure 2.a [Bottom]. In the deformation test, impedance through the TTEs and substrate was measured while a pre-set deformation protocol was applied on the TTE/substrate assembly. Notably, The rubber substrate has elasticity in the order of magnitude of the human skin's one (9.8 MPa vs. 4-20 MPa of skin [44]) and variation of resistance (due to piezoresistive effect) with the applied stretching (0-10%) in the range of the impedance change caused by the respiration (i.e. in the order of 0.1-1 $\Omega$ @64kHz), thus providing a valid test platform for the specific evaluation.

As reported in the experimental section, we defined two types of tests, named respectively "short term" and "long term" stretching tests. The first ones were designed to evaluate the range of stretchability of the TTEs and the interconnections, applying to the TTE/substrate assembly a uniaxial tensile deformation up to 10%. This value has been chosen as maximum elongation for our specific application, since larger of typical maximum stretching which the skin on the chest is subjected during a high inspiration act [45]. Ten TTEs were tested in short term stretching conditions, and all the tests were successful, i.e. the tattoo did not break or crack during the stretching, and the bio-impedance signal acquired by the tattoo did not show discontinuities or non-linearity as a consequence of the applied deformation up to 10%. An example of the results on a typical sample is reported in Figure 2.b, showing the relation between impedance variation and substrate deformation.

"Long term stretching test" were designed for the evaluation of the durability: a cyclic uniaxial tensile deformation was applied to the TTE/substrate assembly up to 96 hours, using a stretching amplitude and rate which simulate the typical respiratory process for a healthy adult, i.e. 3% and 18 cycles/min, respectively. We decided to stop the tests after 96 hours even though the samples were still working as this value greatly overestimates the minimum target value for the proposed application (24 h - TTEs are designed to be replaced daily). Figure 2.c shows an example of the variation of the impedance along time. Moreover, during the long-term stretching test, a vertical acceleration ("shaking") to mimic walking/running movements was applied at constant time intervals, and the quality of the signal and the stability of interconnections during the shaking was evaluated. Figure 2.d shows an example of the impedance signal during 2 minutes of shaking: the curve shows some signal perturbations due to the vertical movements, but the quality of the signal is still more than suitable for the specific application (see also Supplementary Video SV1 for an overview of running experimental setup).

Ten TTEs were tested in long term stretching conditions. The results of long-term stretching test are summarized in Figure 2.e, that shows the distribution of the tattoo failure time. Overall, 70% of the samples were able to achieve 96 hours of stretching. Premature failure of 3 samples was caused by the formation of cracks along the Ag tracks, in apparently random locations. This may be due to the presence of small defects in the tracks, caused by the manual screen printing fabrication process. The use of a professional screen printer could improve the quality of the printing, and consequently the durability of the tracks. Nevertheless, all the samples reached at least 48 hours of stretching, i.e. twice the minimum duration defined for the specific application (24 hours).

The same long-term tests have been also performed on aged samples, to demonstrate the stability of the TTE over time in standard storage condition. In particular, a batch of 10 TTE samples has been fabricated and stored in air at room temperature for 6 months. The batch has been then tested with the long term stretching test protocol, obtaining results that are comparable with the ones obtained with fresh samples, as reported in Figure 2.f (70% of samples lasted more than 96 hours; all samples lasted more than 24 hours).

### 3.3 Test on subject

TTEs for transthoracic impedance measurements were tested on a volunteer subject, chosen among the authors, to evaluate the ability to acquire a reliable signal for the real-time monitoring of patient respiration in real life activities. The volunteer performed a series of standardised routines simulating the daily human real life activities (see detailed protocol in Materials and Methods), while acquiring the transthoracic impedance through the tattoo electrodes and in parallel, as control, the breathing through a thermistor placed close the nose. Both devices are connected with the shimmer unit for real time acquisition/logging of data (see Figure 3.a).

Measurement are analysed off-line. Both bio-impedance signal and thermistor signal are filtered off-line with a Butterworth zero-phase digital filter. The thermistor signal works as a reference signal, being theoretically immune to movement artefacts, even if some ripples in the signal can still occur because of complex mechanisms in the respiratory phase. Nevertheless, they are considered not relevant for the current study. The local maxima and then the onsets of the respiration are determined manually from the two filtered signals: features such as the minimum respiration cycle length and amplitude of adjacent local maxima, are used to this intent. Finally, the detected onsets in the reference and bio-impedance signals are compared. One onset in the reference signal is considered successfully detected, if the correspondent onset in the bio-impedance signal is detected within a time windows of ±500 ms. The detection accuracy is defined as the number of true positive onsets that are detected with the bio-impedance, divided by the total number of true respiration onsets in the reference signal.

An example of the results obtained acquiring in parallel impedance variation and breathing is reported in Figure 3.b. The average detection accuracy on the volunteer subject over the different manoeuvres performed is approximately equal to 92%.

Volunteer user which worn the tattoo on skin was asked to describe whether he perceived or not the presence of tattoo and if his movements were affected or hampered by tattoo. The user described not to perceive the tattoo and did not describe any constraint of his natural movements. After the use, the temporary tattoo electrodes could be removed by washing with soap and water and gently rubbing. The use of commercial medical adhesive remover facilitates the removal of the tattoo.

## 3.4 Other use-case examples, with miniaturised standalone device

The proposed approach can be successfully exploited for a multitude of different applications, including among others health monitoring systems and human machine interfaces.

To explore these possibilities, we developed a small open platform that can be magnetically docked on skin-transferred tattoo electrode and is able to transmit measured data over Bluetooth (BT) connection, thus demonstrating the approach for specific use in EMG recording with threshold detection and in ECG signal monitoring (see Figure 4). The device is built in a modular way to easily adapt to different types of measurement, having a BT microcontroller board for data collection/elaboration and transmission, and a different sensor board implementing the analog front-end for each specific application. Details on the device fabrication, electronic schemes, firmware and software details are reported in Supporting Information and Open Source available (see SI - Stand-alone device for Tattoo Electrode interface – HW details). Release of tattoo and operating Bluetooth module docking/connection is also shown in Supporting Video SV2.

### 3.4.1 EMG measurement and threshold detection

As a first supplementary demonstrative use case we investigated the possibility to record a filtered EMG signal and to reliably detect fixed signal threshold. For EMG detection we used a different tattoo design, as shown in Figure 4.b. In this case, three electrodes are needed, a pair for detecting the differential signal generated by muscle contraction, the third one as reference. The analog front-end used for signal conditioning is schematically reported in Figure 4.a, and it is composed by a first differential amplification stage with a gain $G_1 = 10$, followed by a rectification stage, an active filter module and finally by a second amplification stage with tuneable gain. The output signal is acquired by BT microcontroller board with 12 bit of resolution (full span 3V), further elaborated (the derivative signal is calculated at 50 Hz update rate) and transmitted via BT with a UART (Universal asynchronous receiver-transmitter) over BT protocol to a PC where it is visualised in real time by means of a custom software interface (see in Supporting Information for details). The EMG device has been tested on the arm of a volunteer chosen among authors; the three electrodes tattoo was placed on the forearm in a position suitable to detect the contraction of the wrist flexor muscles, activated by the wrist flexion and hand closure (see Figure 4.c). Once the EMG BT device was connected magnetically to the tattoo, the signal was correctly acquired and transmitted to the PC to show and store the data. When the muscles were relaxed very low background signal was detected (in the order of few mV), while contracting the muscles a very clear and high signal was reported (up to 1 V with built in hardware amplification). Noteworthy strongly shaking the forearm without contracting the muscles resulted in a low background signal, comparable with the one produced at rest (see Figure 4.c and Supporting Video SV3), thus demonstrating that the system (thanks to the magneto-electric connections and smart data acquisition) is quite robust against motion artefacts, also in "real world" conditions. The EMG BT device could find application as muscle-controlled wireless interface in several interesting scenarios, including prosthetic control [12] or interface for entertainment and gaming. As an example, muscle activated remote control of toy car was also demonstrated (see Supporting Video SV4).

### 3.4.1 ECG recording

We also investigated the possibility to measure and transmit ECG signal with stand-alone device. For ECG detection we used a different tattoo design, as shown in Figure 4.e, consisting of two bigger electrodes (2.5 cm of diameter), printed at about 10 cm distance each other. The tattoo electrode design has been optimised to be applied on the chest of the subject to acquire the signal correctly. In this case, two electrodes are sufficient, since the analog front-end chosen for the specific application has a built in reference system. In more details, the analog module is different from the previous one presented for ECG (while the BT microcontroller board is the same, see Figure 4.d, and it is specifically built around the MAX30003 chip (Maxim Integrated, U.S.), a monolithic biopotential, analog front-end for clinical and fitness applications, providing ECG waveforms and heart rate detection. The MAX30003 is connected with the microcontroller through a high speed digital SPI interface, guaranteeing up to 500 sps (128 sps in our specific implementation) of waveform sampling with a resolution up to 15 bit and $5\mu V_{P-P}$ noise.

The ECG device has been tested on the chest of a volunteer chosen among authors; the two-electrodes tattoo was placed at in the middle of the chest in correspondence of the sternum. A comparative test was then performed by using as reference a commercial handheld device (Prince 180B Easy ECG Monitor, Heal Force Bio-meditech, Shanghai, China) equipped with three standard pre-gelled electrodes: the two signal electrodes were placed just below the ECG tattoo electrodes (see Figure 4.e), while the ground reference one was placed far from the site measurement, on a leg. Once the ECG BT device was magnetically connected to the tattoo, synchronised acquisition was started. Several acquisitions lasting each 60 s each were performed, and tattoo signal compared with reference one (see Figure 4.e bottom for typical measurement results, see also Supporting Video SV5). In all the trials the signals acquired (tattoo electrodes and gelled references) are mostly superimposable (apart for a scale factor), thus demonstrating the perfect functioning of the device. It is also worth to underline that the resolution of the ECG BT device (<1μV) resulted superior respect to the commercial device one (10μV), making possible the detection of smaller features of the waveform. Also in term of RMS noise and signal to noise ratio the Tattoo ECG BT system is over performing the reference one. In fact, estimating the "noise" RMS amplitude on a full 60 second sample, as the square root of the integral of the power spectrum (at mean squared amplitude, MSA) above 40 Hz [14], the amplitude resulted to be 0.29 μV with tattoo ECG BT system and 2.00 μV with reference device. The total RMS amplitude was instead 62.96 μV versus 156.70 μV, giving a signal-to-noise ratio of 211 and 78, respectively (see also Supporting Materials SI.8). Specifically, the EMG BT device could easily find application as low cost, high accuracy personal tool for heart monitoring.

## 4. Conclusions

In this work, a new strategy for the establishment of easy-to-handle/reliable and long-term stable interconnections between ultra-conformable temporary tattoo electrodes (TTEs) and external magnetically docked devices has been presented. The proposed approach was validated on a specific real world use case with clinical relevance, i.e. a disposable epidermal sensor for real time monitoring of respiration. Moreover, some proof-of-concept demonstrators were presented, further demonstrating the approach for specific use in EMG recording with threshold detection and in ECG signal monitoring. The efficiency of the TTE and the proposed approach under stretching (up to 10%) and over time (up to 96 hours) has been verified on a dedicated experimental setup and further demonstrated with a dedicated test on human subject. The proposed design make TTE technology amenable for large-scale production of low-cost skin-contact sensing devices which could be successfully exploited for a multitude of different applications, including among others health monitoring systems and human machine interfaces.

**Experimental**

**- Tattoo fabrication**

Commercially available temporary transfer tattoo paper kit (Silhouette Tattoo Paper, Silhouette America, USA) composed of two sheets (a decal transfer paper and a glue sheet) was employed The surface of the decal transfer paper, used as unconventional substrate for electrodes fabrication, has been gently washed with a DI water jet and then dried using a compressed-air gun. In order to avoid its wetting, the back of the paper was covered with an aluminum foil and edges protected by impermeable adhesive tape. Tattoo paper outline and central hole have been cut by using a $CO_2$ laser cutter (model VLS3.50 by Universal Laser System). PEDOT:PSS electrodes have been deposited by inkjet printing of a solution of PEDOT:PSS aqueous dispersion (Clevios PJet 700 by Heraeus) and glycerol (10% vol). PEDOT:PSS ink was used after filtration (Minisart, average pore size 0.20 μm, Sartorius). Inkjet printing was carried out with a Dimatix DMP-2800 system (Fujifilm Corp., Japan) endowed with a 10 pL cartridge (DMC-11610). Interconnection Ag tracks on tattoo paper have been deposited by screen printing of a stretchable Ag conductor paste (CI-1036, Engineered Materials Systems). For the screen printing, we used a homemade setup for manual screen printing, composed of a mask (laser cut stencil) and a blade for squeegeeing and uniformly distribute the paste. After the deposition, a baking at 120°C for 10 min was performed. A 25 μm thick polyimide foil (Kapton HN, acquired from RS) was employed as support layer for the external electrical connection (contact pad). The outline of the contact pad has been cut using a $CO_2$ laser cutter (model VLS3.50 by Universal Laser System). Silver pads on contact pad have been deposited by screen printing of the stretchable Ag conductor paste through a stencil mask. After the deposition, a baking at 120°C for 10 min was performed. Contact pads were flipped for correct assembly. During the assembly of the two parts, the Contact pad was glued to the tattoo putting a small drop of the Ag paste between the tracks and the pads, and then baked at 120°C for 10 min. The glue sheet of the temporary transfer tattoo paper kit was firstly laser cut, and then placed on top of the device for providing additional tattoo substrate adhesion (except for the sensing PEDOT:PSS parts) while acting as protective insulating layer, preventing direct contact of interconnection lines with skin. Four small neodymium magnets (0.5 mm thick, 2.5 mm diameter) were fixed with glue on a plastic support (0.5 mm thick) to obtain a magnetic disk. The magnetic disk was finally attached to the on-tattoo mounted contact pad, on the opposite side of the silver pad obtaining the final sensor assembly ready for use.

**- Tattoo characterization**

Surface analysis was performed on tattoo samples, recollected onto different specific supports, after the release in water as free standing membrane. Thickness measurements were carried out with a P6 stylus profilometer, KLA Tencor, onto samples recollected on clean a Si wafer.

**- In-lab stretching test to evaluate the connection reliability**

Stretching tests were carried out using a home-made measurement platform. Conductive rubber samples (Axel 2-9330-02) were fixed at one edge and clamped to a motor driven translation stage on the other (L-509, Physik Instrumente) in order to apply a uniaxial tensile deformation along the transverse axis. The control and conditioning electronics related to the translation and testing stage was connected through a multichannel DAQ Board (model USB-6218, National Instruments, US) with a dedicated PC. Stage position control was performed using a customized Graphical User

Interface (GUI) developed with Visual Studio 2010 (Microsoft, Redmond, CA). The setup for the stretch has been fixed on a vertical plate which can move along y axis. A gearmotor (100:1 Metal Gearmotor 37Dx73L mm with 64 CPR Encoder) has been employed for vertical movement. Short-term stretching test, deformation applied: 3 – 4 – 5 – 6 – 7 – 8 – 9 – 10% (every deformation repeated 5 times, stretching speed was fixed to about 2 % $s^{-1}$). Long-term stretching test, deformation applied: 3% with a frequency of 18 cycles/min. During long term stretching test, a vertical vibration (acceleration range 4g) of 2 min was applied at constant time intervals of 1 hour.

TTE was transferred on fixed conductive rubber just before the start of experiments, by gently wetting the backside of the TTE sample placed on the rubber surface with a wetted sponged for removing the paper support. Impedance measurements were performed by using a Shimmer 3 ECG wireless sensor unit (Shimmer, Ireland), configured for 4-point impedance measurement, with excitation AC current of 30 μA @64kHz, and sampling rate of 300 Hz for short-term measurements and 1 Hz for long-term ones. The shimmer unit was electrically connected with the TTE trough a wired magnetic docking pad (as shown in Figure 2.a).

- Tests on volunteer - protocol description

A volunteer, chose among the authors, performed a series of standardised routines simulating the daily human real life activities while monitoring simultaneously the transthoracic impedance and the breathing through a thermistor placed close the nose. The volunteer is prepared as reported in the following steps: 1) The skin of the participant is gently rubbed with an antibacterial solution to remove the natural dermal fat coating; 2) The TTE is applied in correspondence of the volunteer's sternum, at the Louis angle; 3) A NTC thermistor (Alice 6 respiratory sensor nasal/oral thermistor, Ternimed UG, Germany) is placed in the patient's nose to provide breathing reference signal; 4) The Shimmer 3 ECG wireless sensor unit (Shimmer, Ireland), configured for 4 point impedance measurement, (excitation AC current of 30 μA @64kHz, and sampling rate of 256 Hz), is placed right up the patch by using a double layer adhesive tape; 5) The Shimmer3 is wired-connected to the thermocouple and to the TTE's though a magnetically docking cables.

The measurement protocol used in this investigation emulates the daily human activities, including six manoeuvres that represent the typical actions performed during a normal 24 hours' real life day. In details, the 7 minutes' measurement protocol includes: short-time breathing holding (up to patient ability/up to max 30 s); normal breathing (60 s); loud voice reading (60 s); left/right head rotation (30 s); left/right torso rotation and inclination ahead/back (30 s); slow paced walking (2 km/h; 60 s); medium paced walking (3 km/h; 60 s); free talking (60 s). Data for transthoracic impedance and breathing are acquired by Shimmer3 during all the protocol and analysed offline.

**Conflict of Interest and Sources of Funding:** The development of the TTe was sponsored by MED-EL Elektromedizinische Geräte GmbH, Innsbruck, Austria.

**References**

1. D. H. Kim, N. S. Lu, R. Ma, Y. S. Kim, R. H. Kim, S. D. Wang, J. Wu, S. M. Won, H. Tao, A. Islam, K. J. Yu, T. I. Kim, R. Chowdhury, M. Ying, L. Xu, M. Li, H. J. Chung, H. Keum, M. McCormick, P. Liu, Y. W. Zhang, F. G. Omenetto, Y. G. Huang, T. Coleman, J. A. Rogers, *Science* **2011**, *333*, 838.

2. M. L. Hammock, A. Chortos, B. C. Tee, J. B. Tok, Z. Bao, *Adv. Mater.* **2013**, *25*, 5997.

3. Y. Liu, M. Pharr, G. A. Salvatore, *ACS Nano* **2017**, *11*, 9614.

4. B. Xu, A. Akhtar, Y. Liu, H. Chen, W. H. Yeo, S. I. Park, B. Boyce, H. Kim, J. Yu, H. Y. Lai, S. Jung, Y. Zhou, J. Kim, S. Cho, Y. Huang, T. Bretl, J. A. Rogers, *Adv. Mater.* **2016**, *28*, 4462.

5. J. W. Jeong, W. H. Yeo, A. Akhtar, J. J. Norton, Y. J. Kwack, S. Li, S. Y. Jung, Y. Su, W. Lee, J. Xia, H. Cheng, Y. Huang, W.-S. Choi, T. Bretl, J. A. Rogers, *Adv. Mater.* **2013**, *25*, 6839.

6. S. Wang, M. Li, J. Wu, D.-H. Kim, N. Lu, Y. Su, Z. Kang, Y. Huang, J. A. Rogers, *J. Appl. Mech.* **2012**, *79*, 031022.

7. A. Zucca, K. Yamagishi, T. Fujie, S. Takeoka, V. Mattoli, F. Greco *J. Mater. Chem. C* **2015**, *3*, 6539.

8. K. Yamagishi, S. Taccola, S. Takeoka, T. Fujie, V. Mattoli, F. Greco, *Flexible and Stretchable Medical Devices*, Wiley-VCH Verlag GmbH & Co. KGaA, Weinheim, **2018**.

9. L. Wang, N. Lu, *J. Appl. Mech.* **2016**, *83*, 041007.

10. J. W. Jeong, M. K. Kim, H. Cheng, W. H. Yeo, X. Huang, Y. Liu, Y. Zhang, Y. Huang, J. A. Rogers, *Adv. Health. Mater.* **2014**, *3*, 642.

11. Y. Wang, Y. Qiu, S. K. Ameri, H. Jang, Z. Dai, Y. Huang, N. Lu, *NPJ. Flexible Electron.* **2018**, *2*, 6.

12. L. M. Ferrari, K. Keller, B. Burtscher, F. Greco Multifunct. Mater **2020**. *3*, 032003.

13. A. Zucca, C. Cipriani, Sudha, S. Tarantino, D. Ricci, V. Mattoli, F. Greco, *Adv. Healthcare Mater.* **2015**, *4*, 983.

14. L. M. Ferrari, S. Sudha, S. Tarantino, R. Esposti, F. Bolzoni, P. Cavallari, C. Cipriani, V. Mattoli, F. Greco, *Adv. Sci.* **2018**, *5*, 1700771.

15. A. Chortos, G. I. Koleilat, R. Pfattner, D. Kong, P. Lin, R. Nur, T. Lei, H. Wang, N. Liu, Y. C. Lai, M. G. Kim, J. W. Chung, S. Lee, Z. Bao, *Adv. Mater.* **2016**, *28*, 4441.

16. J. Kim, W. R. de Araujo, I. A. Samek, A. J. Bandodkar, W. Jia, B. Brunetti, T. R. L. C. Paixão, J. Wang, *Electrochem. Commun.* **2015**, *51*, 41.

17. A. J. Bandodkar, W. Jia, C. Yardimci, X. Wang, J. Ramirez, J. Wang, *Anal. Chem.* **2015**, *87*, 394.

18. L. Bareket, L. Inzelberg, D. Rand, M. David-Pur, D. Rabinovich, B. Brandes, Y. Hanein, *Sci. Rep.* **2016**, *6*, 25727.

19. E. Bihar, T. Roberts, Y. Zhang, E. Ismailova1, T. Hervé, G. G. Malliaras, J. B. De Graaf, S. Inal, M. Saadaoui, *Flex. Print. Electron.* **2018**, *3*, 034004.

20. G. E. Bonacchini, C. Bossio, F. Greco, V. Mattoli, Y.-H. Kim, G. Lanzani, M. Caironi, *Adv. Mater.* **2018**, *30*, 1706091.
21. N. Piva, F. Greco, M. Garbugli, A. Iacchetti, V. Mattoli, M. Caironi, *Adv. Electron. Mater.* **2018**, *4*, 1700325.
22. P. Leleux, C. Johnson, X. Strakosas, J. Rivnay, T. Herve, R. M. Owens, G. G. Malliaras, *Adv. Healthcare Mater.* **2014**, *3*, 1377.
23. L. M. Ferrari, U. Ismailov, J.-M. Badier, F. Greco, E. Ismailova, *npj Flexible Electronics* **2020**, *4*, 4.
24. S. Wagner, S. Bauer, MRS Bull. **2012**, *37*, 207.
25. S. Gong, L. W. Yap, B. Zhu, W. Cheng, *Adv. Mater.* **2020**, *32*, 1902278.
26. S. Xu, Z. Yan, K.-I. Jang, W. Huang, H. Fu, J. Kim, Z. Wei, M. Flavin, J. McCracken, R. Wang, A. Badea, Y. Liu, D. Xiao, G. Zhou, J. Lee, H. U. Chung, H. Cheng, W. Ren, A. Banks, X. Li, U. Paik, R. G. Nuzzo, Y. Huang, Z. Yihui, *Science* **2015**, *347*, 154.
27. Y. Sun, W. M. Choi, H. Jiang, Y. Y. Huang, J. A. Rogers, *Nat. Nanotechnol.* **2006**, *1*, 201.
28. Y. Zhang, S. Wang, X. Li, J. A. Fan, S. Xu, Y. M. Song, K. J. Choi, W. H. Yeo, W. Lee, S. N. Nazaar, Bingwei Lu, L. Yin, K.-C. Hwang, J. A. Rogers, Y. Huang, *Adv. Funct. Mater.* **2014**, *24*, 2028.
29. Y. Ling, K. Guo, B. Zhu, B. Prieto-Simon, N. H. Voelcker, W. Cheng, *Nanoscale Horiz.* **2019**, *4*, 1380-1387.
30. M. A. Cretikos, R. Bellomo, K. Hillman, J. Chen, S. Finfer, A. Flabouris, *Med. J. Aust.* **2008**, *188*, 657-659.
31. Massaroni, A. Nicolò, D. Lo Presti, M. Sacchetti, S. Silvestri, E. Schena, *Sensors* **2019**, *19*, 908.
32. E. Vanegas, R. Igual, I. Plaza, *Sensors* **2020**, *20*, 5446.
33. M. Wang, J. Zhang, Y. Tang, J. Li, B. Zhang, E. Liang, Y. Mao, X. Wang, *ACS Nano* **2018**, *12*, 6156−6162.
34. H. Jin, X. Tao, S. Dong, Y. Qin, L. Yu, J. Luo, M. J. Deen, *J. Micromech. Microeng.* **2017**, *27*, 115006.
35. Y. Pang, J. Jian, T. Tu, Z. Yang, J. Ling, Y. Li, X. Wang, Y. Qiao, H. Tian, Y. Yang, T.-L. Ren, *Biosensors and Bioelectronics* **2018**, *116*, 123–129.
36. M. Chu, T. Nguyen, V. Pandey, Y. Zhou, H. N. Pham, R. Bar-Yoseph, S. Radom-Aizik, R. Jain, D. M. Cooper, M. Khine, *npj Digital Medicine* **2019**, 2, 8.
37. D. Blanco-Almazán, W. Groenendaal, F. Catthoor, R. Jané, *IEEE Access* **2019**, *7*, 89487-89496.
38. T. Dinh, T. Nguyen, H.-P. Phan, N.-T. Nguyen, D. V. Dao, J. Bell, *Biosensors and Bioelectronics* **2020**, *166*, 112460.
39. D. Blanco-Almazán, W. Groenendaal, F. Catthoor, R. Jané, *Sci Rep* **2019**, *9*, 20232.
40. S. Timoshenko, S. Woinowsky-Krieger, *Theory of Plates and Shells*, McGraw-Hill, New York, **1987**.

41. K.-I. Jang, H. N. Jung, J. W. Lee, S. Xu, Y. H. Liu, Y. Ma, J.-W. Jeong, Y. M. Song, J. Kim, B. H. Kim, A. Banks, J. W. Kwak, Y. Yang, D. Shi, Z. Wei, X. Feng, U. Paik, Y. Huang, R. Ghaffari, J. A. Rogers, *Adv Funct Mater.* **2016**, 26(40), 7281-7290.

42. A. K. Gupta, Application Report SBAA181; Texas Instruments: Dallas, TX, USA, **2011**.

43. S. H. Eom, S. Senthilarasu, P. Uthirakumar, S. C. Yoon, J. Lim, C. Lee, H. S. Lim, J. Lee, S.-H. Lee, *Organic Electronics* **2009**, *10*, 536.

44. C. Pailler-Mattei, S. Bec, H. Zahouani, Med. Eng. Phys. **2008**, 30, 599-606.

45. B. Carlson, *Phys. Ther.* **1973**, *53*, 10–14.

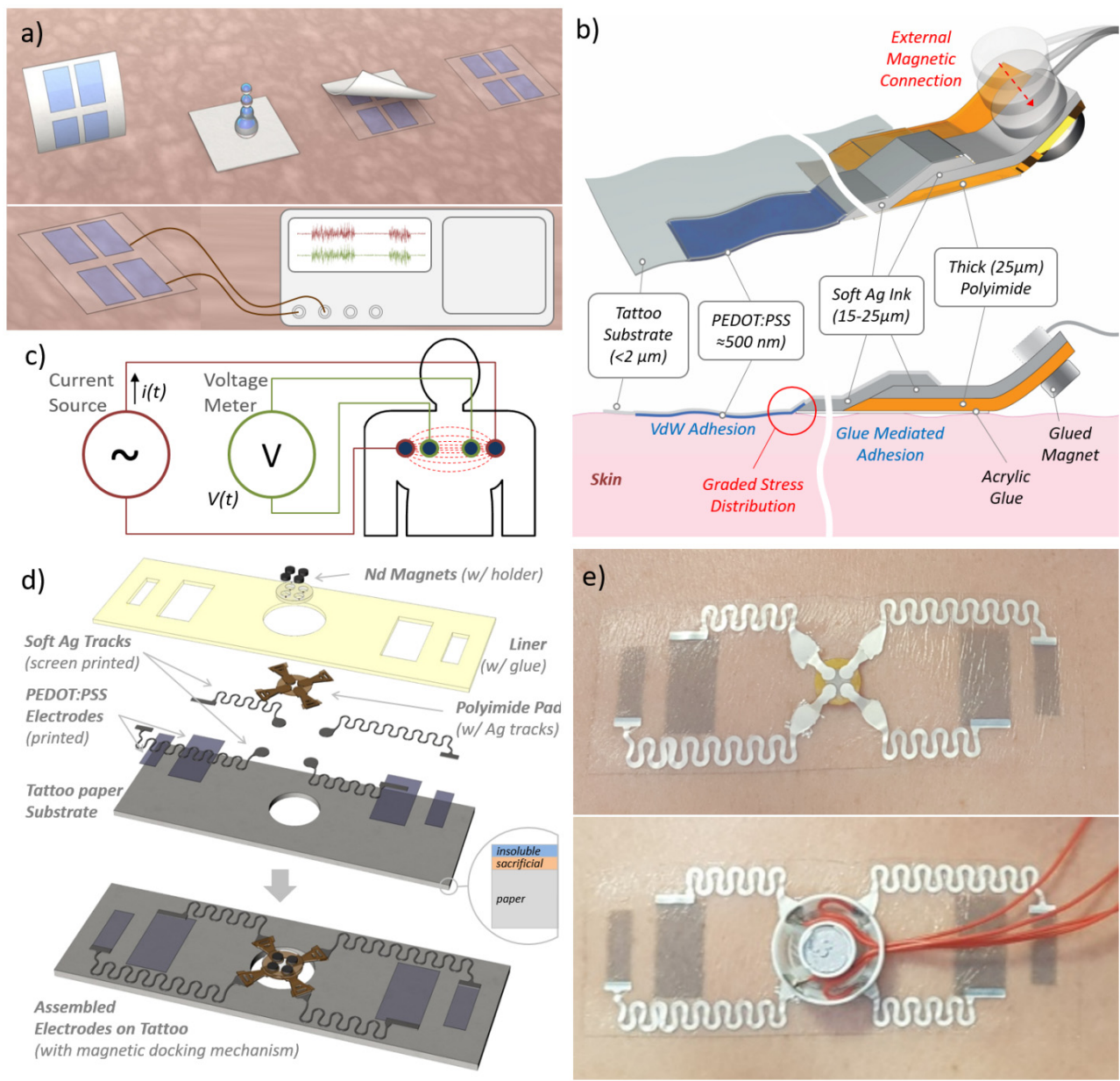

**Figure 1.** a) Schematic representation of transfer on skin of printed temporary tattoo electrodes (up) and wiring to external electronic device (down). Establishing easy-to-handle/reliable and stable interconnections represent a typical challenge for any ultrathin epidermal device; b) Connection schematic section for a single electrode: the use of the soft and stretchable Ag ink allows the interface between materials of different thickness and Young's moduli, i.e. the tattoo electrode (conformally attached to the skin) and the polyimide film. The external magnetic connection allows the communication with the external device; c) The specific case-study: tattoo electrodes as real time respiration sensor based on four-points measurement of bio-impedance; d) Schematic representation of the design and geometry of PEDOT:PSS electrodes and interconnections for transthoracic impedance measurements; e) A TTE released on the chest of a subject (top) and the same docked through a magnetic connector used for signal acquisition.

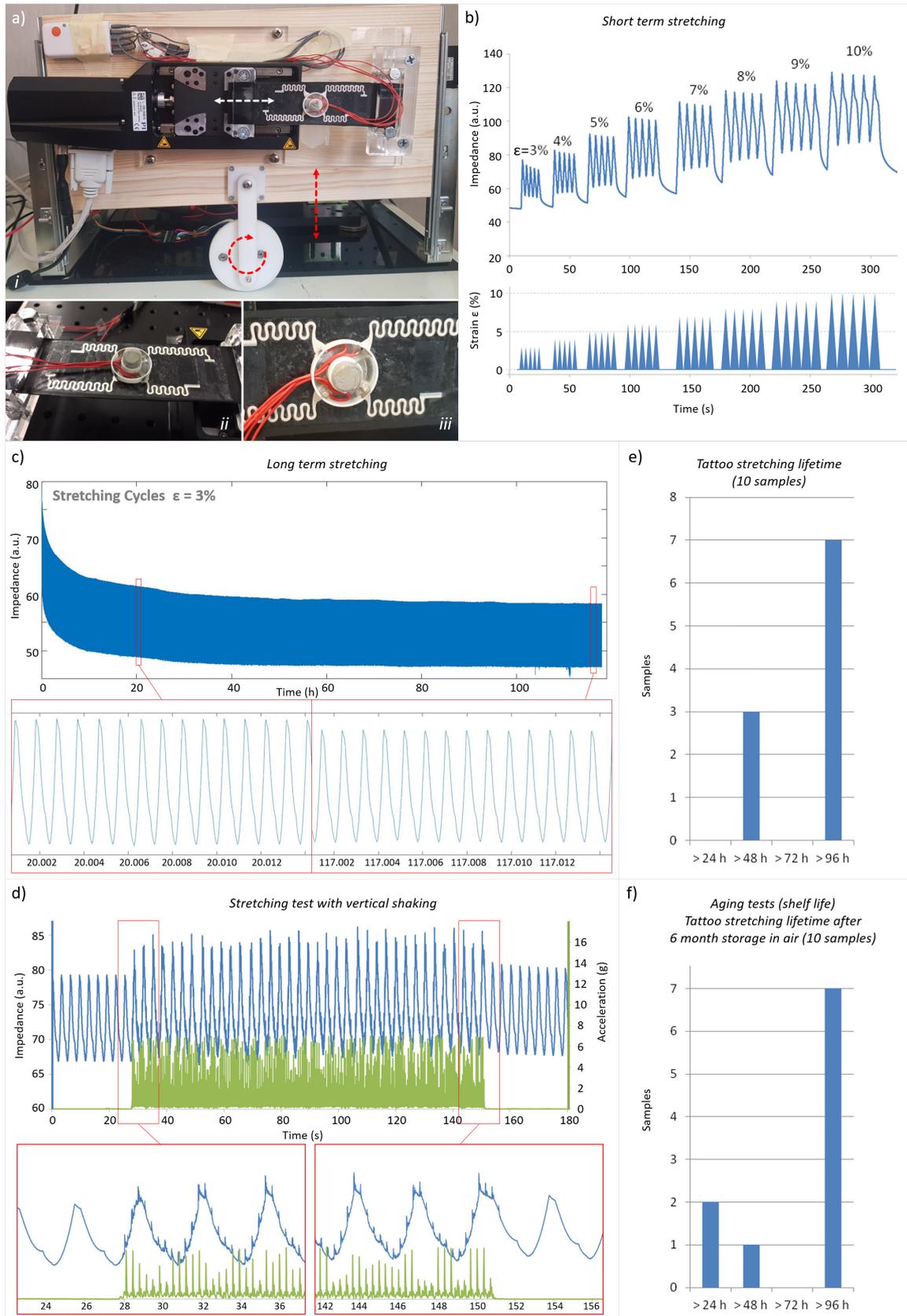

**Figure 2.** a) A picture of the experimental setup used for stretching test (top) and of the TTE transferred onto a conductive and stretchable rubber substrate (bottom); b) "Short term stretching test": an example of the relation between impendence variation measured by the TTE and substrate deformation (up to 10% strain); c) "Long term stretching test": an example of the variation of the impedance along time during a cyclic uniaxial deformation of 3% (frequency 18 cycles/min) up to >96 hours.; d) An example of the impedance signal during 2 minutes of "shaking", showing that the sensor output is scarcely affected by motion artifacts; e) Summary of the results of long term stretching test, showing that 70% of the samples was able to reach the 96 hours of stretching; f) Summary of the results of long term stretching test on 6 months aged samples: stretching lifetime of the tattoo is not significantly deteriorated with aging.

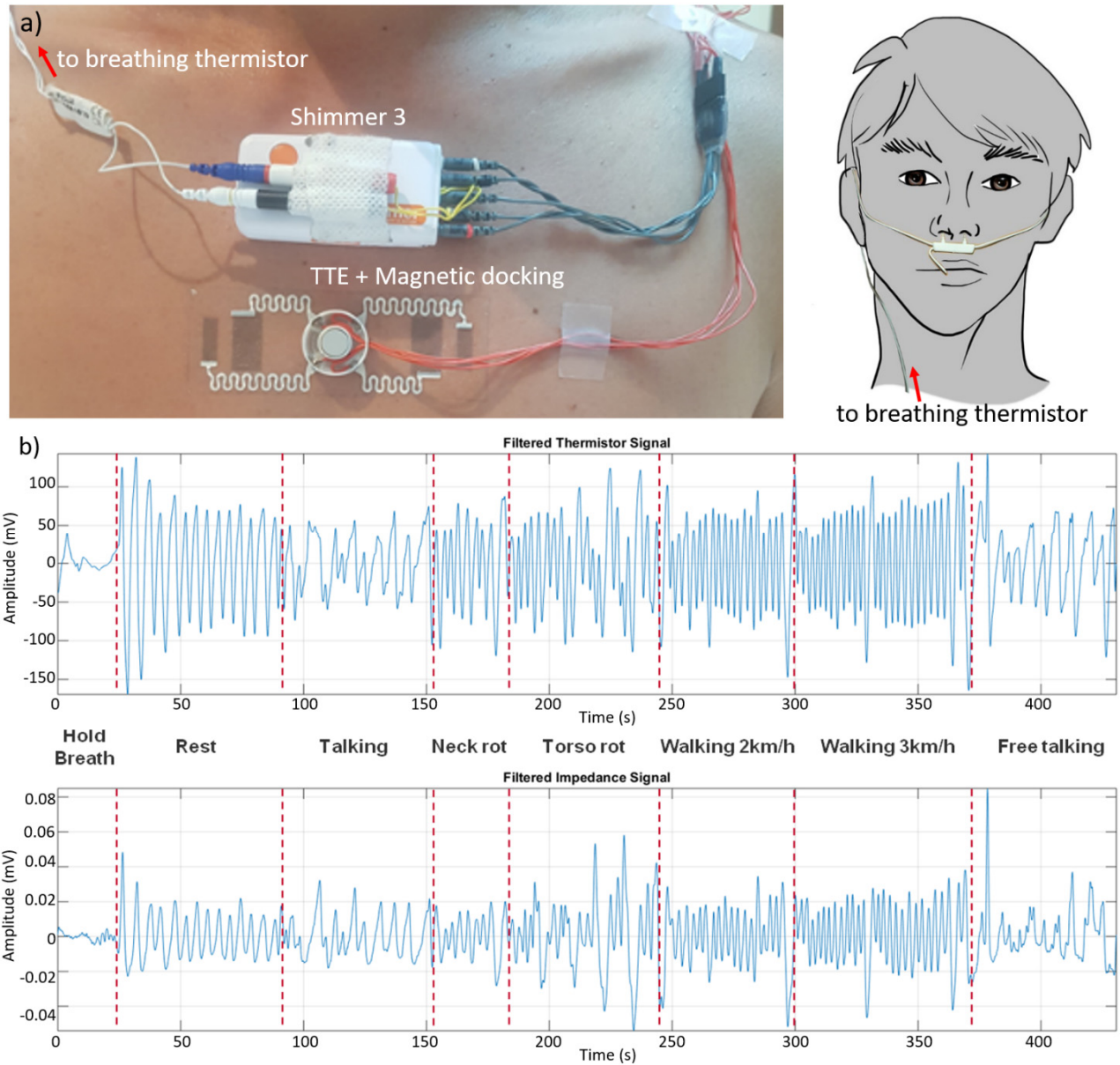

**Figure 3.** a) A picture of a volunteer user wearing TTE on skin during test on humans and interconnections with Shimmer unit used for tests. A breathing thermistor was used for comparison; b) Impedance variation measurements during a series of standardised routines simulating the daily human real life activities (i.e. walking) and comparison between the impedance variation acquired by TTE (top) and breathing measured through a thermistor placed close the nose (bottom).

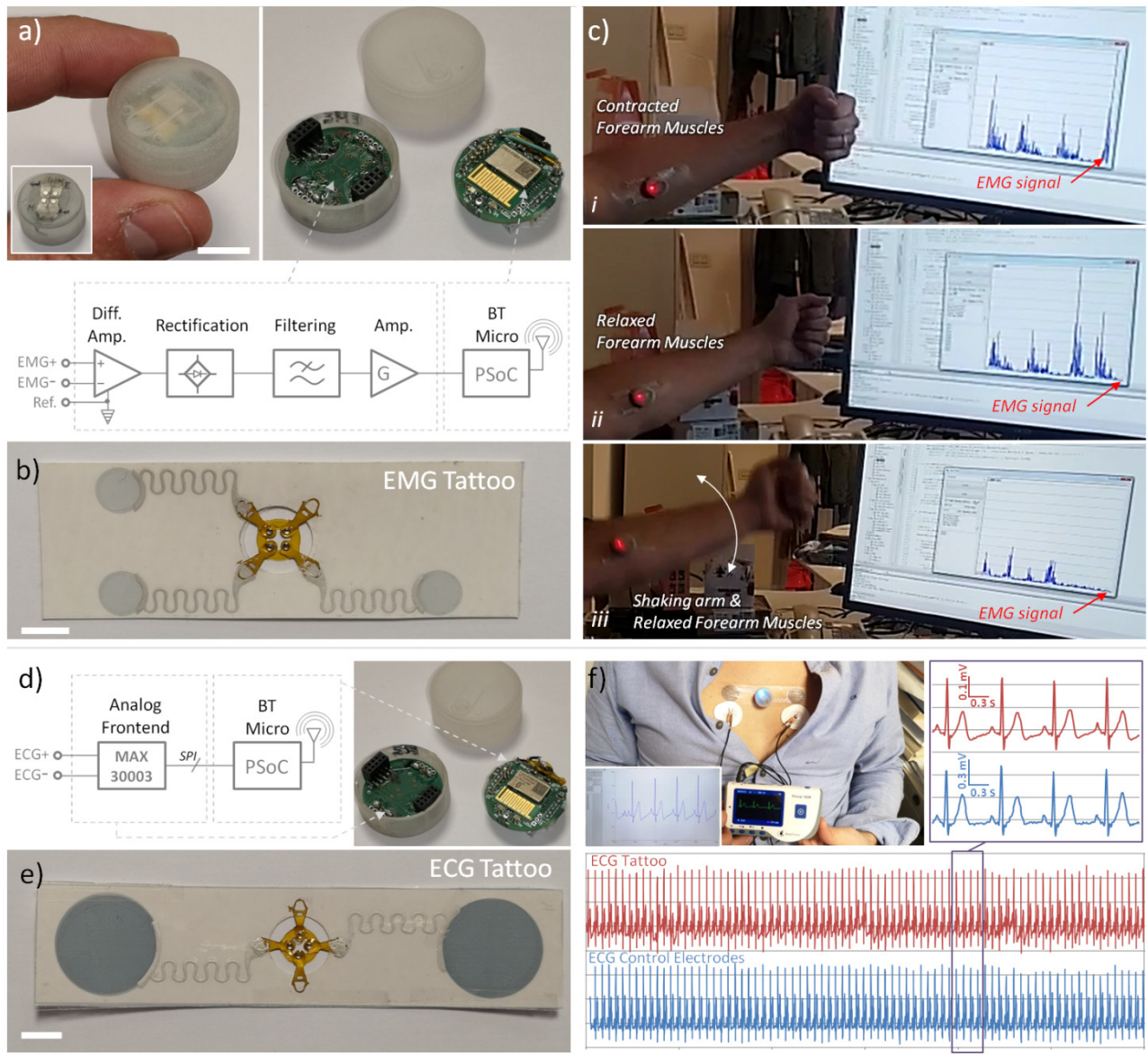

**Figure 4.** a) Picture and schematic of magnetic dockable BT acquisition device, specifically designed to acquire filtered EMG signal (scale bar 1 cm); b) picture three-electrode tattoo for EMG signal acquisition (scale bar 1 cm); c) example of use of tattoo EMG electrodes with BT device transmitting low-pass filtered data to a PC, demonstrating large immunity from motion artefacts: signal is generated exclusively by muscle contraction, not from forearm shaking. d) Picture and schematic of magnetic dockable BT acquisition device, specifically designed to acquire ECG signal; e) picture of two-electrode tattoo for ECG signal acquisition (scale bar 1 cm); f) measurement of ECG carried out by using tattoo ECG electrodes with specific ECG BT device compared with commercial handhold device using three standard pre-gelled electrodes: rescaled graphs are compared in the same time framework; in the inset the zoomed view of a small portion (3 s) of the full acquisition time (60 s) is reported. The matching is almost perfect.

# Toward the use of temporary tattoo electrodes for impedancemetric respiration monitoring and other electrophysiological recordings

*S. Taccola*[1,2], *A. Poliziani*[1], *D. Santonocito*[3], *A. Mondini*[1], *C. Denk*[3], *A. N. Ide*[3], *M. Oberparleiter*[3], *F. Greco*[1,4], *V. Mattoli*[1]

[1]*Center for Micro-BioRobotics, Istituto Italiano di Tecnologia, Viale Rinaldo Piaggio 34, 56025 Pontedera, Pisa, Italy.*

[2] *Future Manufacturing Processes Research Group, School of Mechanical Engineering, Faculty of Engineering, University of Leeds, Leeds LS2 9JT, United Kingdom.*

[3] *Emerging Application Department, MED-EL Elektromedizinische Geräte Gesellschaft m.b.H., Fürstenweg 77a, 6020 Innsbruck, Austria*

[4] *Institute of Solid State Physics, NAWI Graz, Graz University of Technology, Petersgasse 16, 8010 Graz, Austria.*

*\*E-mail: S.Taccola@leeds.ac.uk; francesco.greco@tugraz.at; virgilio.mattoli@iit.it*

**Details on preparation of decal transfer Silhouette paper**

The composition of decal transfer Silhouette paper is (according to available datasheet): paper sheet (contents 83%); polyvinyl alcohol resin (10%); polyamide resin (3%), polyalyl resin (2%) and polyurethane resin (2%).

While the exact composition and arrangement of the different layers is not disclosed in the datasheet, our understanding and investigation make us conclude that Silhouette paper is made up of at least three layers: the paper carrier; a PVA sacrificial water soluble layer (which allows the release of the tattoo), and the tattoo layer which can be transferred onto skin.

Before electrodes and tracks deposition, the surface of the decal transfer paper has been gently washed with a water jet, and then dried using a compressed-air gun. Comparing the morphology of the paper observed by optical microscope before and after the wash, it seems that a water soluble layer on top was removed by washing (Figure SI1).

This hypothesis was confirmed by thickness measurements of the released tattoo layer without or with wash: washing of the tattoo paper sheet prior to release reduced the final thickness of the tattoo layer from about 4-4.5 μm to about 1.7 um.

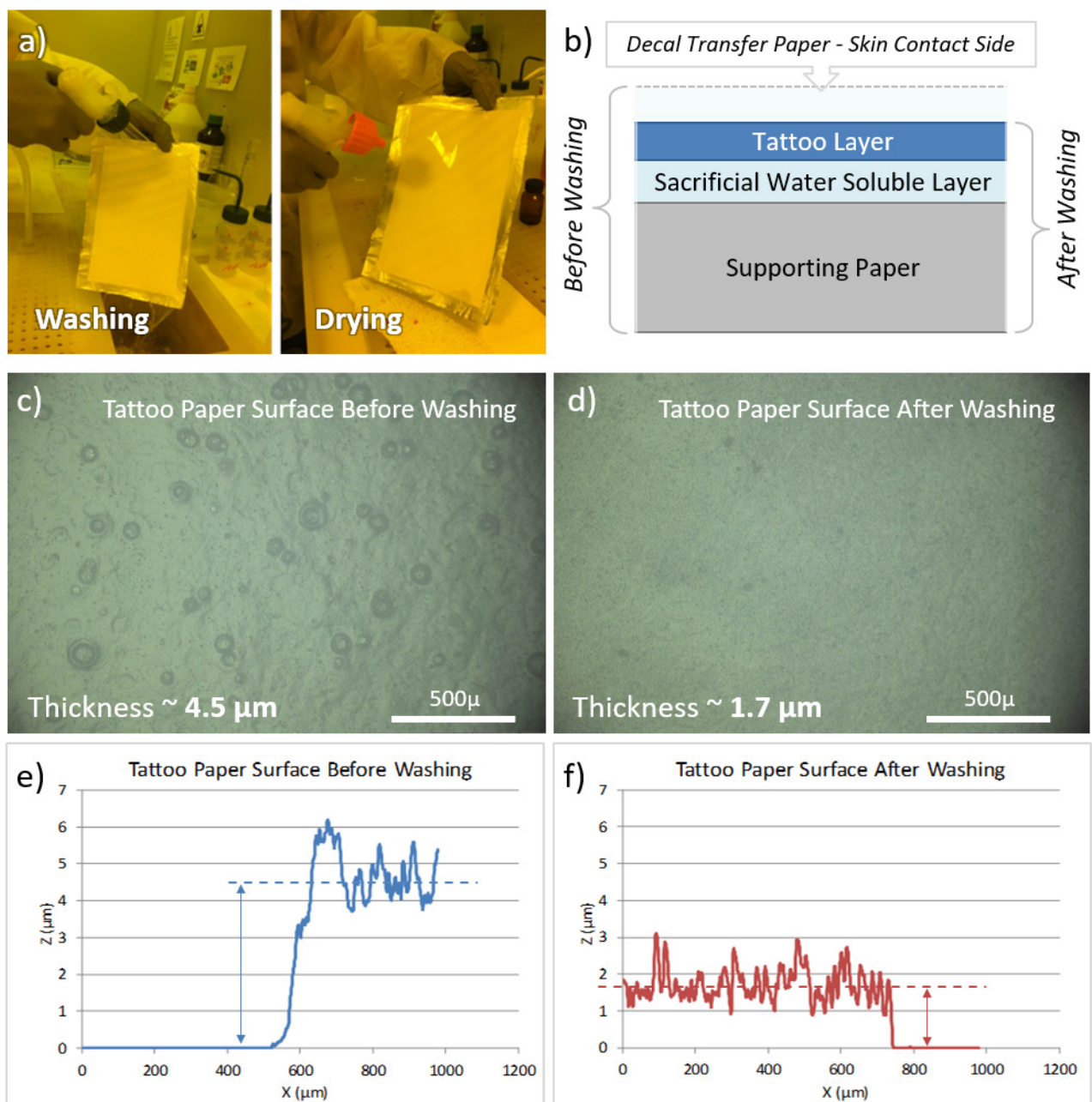

**Figure SI1.** a) Washing and drying of the surface of the decal transfer paper using a DI water jet (left) and a compressed-air gun (right); b) A schematic representation of the different layers which compose the paper before and after washing and the corresponding optical microscope images of the paper surface before (c) and after (d) the treatment; Typical mean value for decal transfer paper thickness before and after washing are reported, togheter with the corresponding examples of thickness profiles (e and f).

## Tattoo electrodes -fabrication procedure

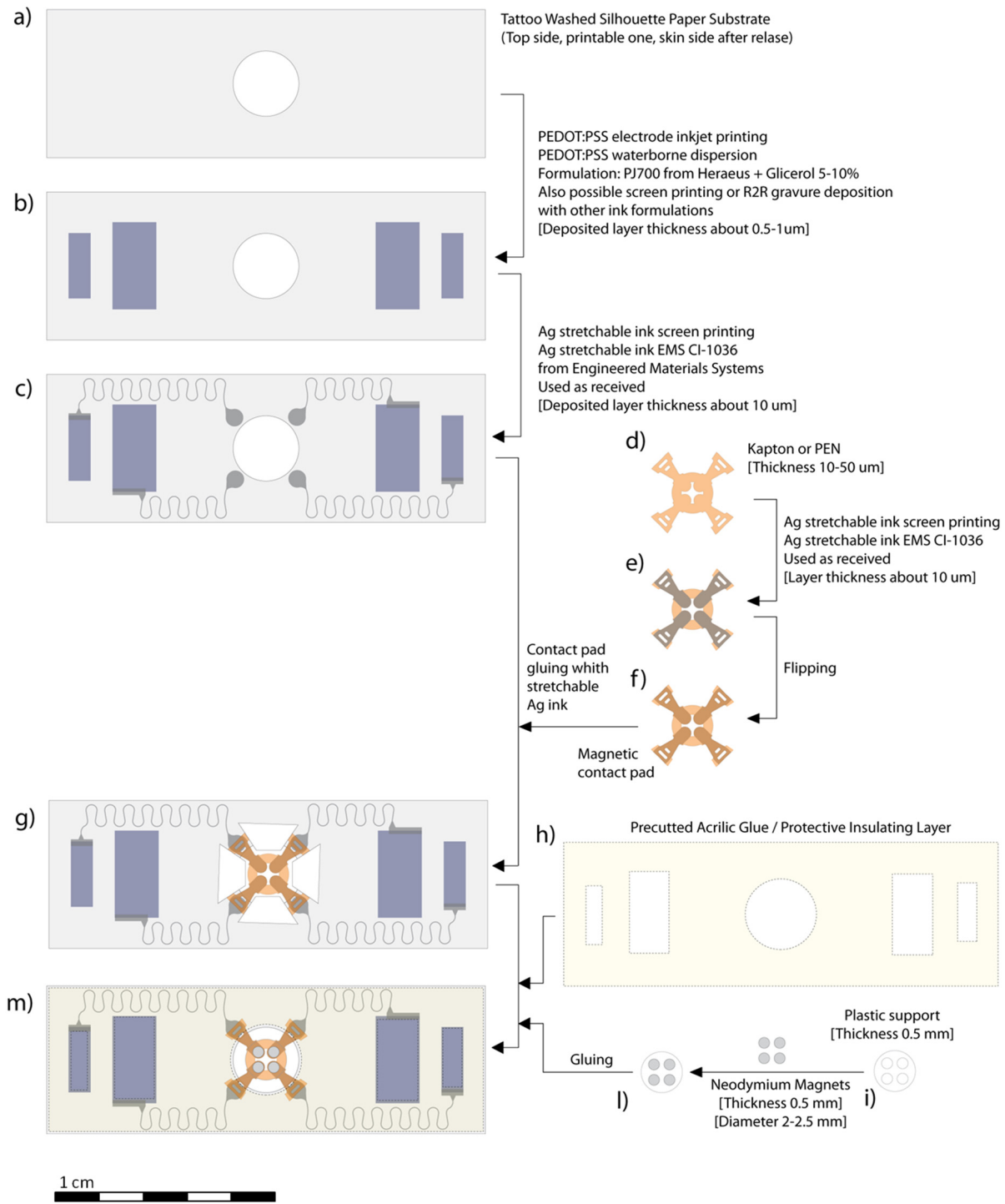

**Figure SI.2.** Schematic representation of Tattoo electrodes fabrication procedure.

## Tattoo electrodes - Inkjet printing and conductivity

In general, concerning ink-jet printing, the properties of the substrate strongly influence the quality of the print, and consequently the conductivity of the printed electrodes. For this reason the printing parameters have to be optimized. In particular we worked on the formulation of PEDOT:PSS ink (conducting polymer in water) using glycerol as a biocompatible additive improving conductivity and print quality (*S.H. Eom et al. Organic Electronics 10, 536–542, 2009*). The final formulation chosen was PEDOT:PSS Clevios P Jet 700 (H.C. Starck) + 10% vol glycerol. In order to print more layers, an intermediate heating at T=120°C for 10 min is needed. In Figure SI.3 it is shown how the addition of glycerol improved the quality of the print and increased the conductivity of PEDOT:PSS electrodes, printed as multiple superimposed layers. "R" is the surface resistivity in Ω/□, measured the opposite edges of a printed square of 1cm lateral size.

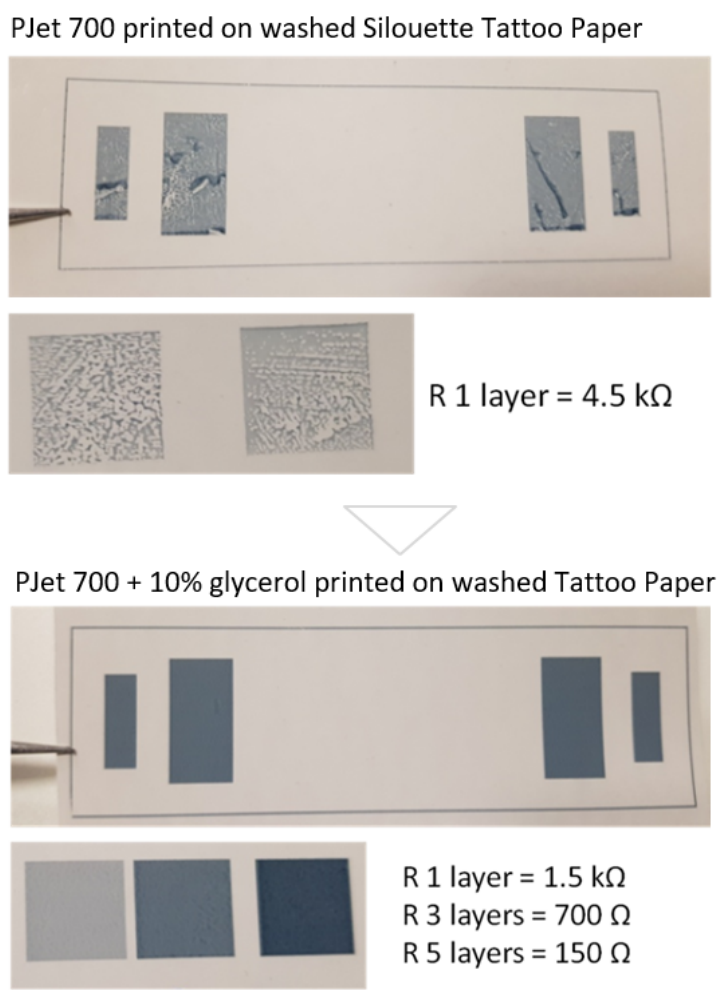

**Figure SI.3.** Example of samples obtained by printing PEDOT:PSS Clevios P Jet 700 as received on washed Silouette tattoo paper (top), and samples obtained by printing PEDOT:PSS Clevios P Jet with addition of 10%$_{vol}$ of glycerol (bottom). Typical value for electrodes' surface resistivity are reported.

**Tattoo electrodes - material's properties**

|  | PEDOT:PSS Layer | Tattoo Substrate | Silver Paste | Kapton |
|---|---|---|---|---|
| Bulk Modulus E [MPa] | $1$-$2 \cdot 10^3$ [a] | 42 [a] | ≈ 100-200 [b] | $2$-$2.7 \cdot 10^3$ [a] |
| Max Strength $S_{max}$ [MPa] | --- | --- | ≈ 10-26 [b] | ≈ 100 [a] |
| Poisson's ratio $\nu$ | 0.3 [c] | 0.5 [c] | 0.5 [c] | 0.34 [a] |
| Max strain $\varepsilon_{max}$ | 5-10% (uniax) [b] | >10% [b] | >10% [b] | <1% [a] |
| Thickness [um] | 0.4-0.6 [b] | 1.5 [b] | 15 [d] | 25 [a] |
| Flexural rigidity D [N m x$10^{-9}$] [e] | 0.0223 | 0.0236 | 56.3 | 4650 |

**Table SI.1**. Summary of typical mechanical properties for used materials.

[a] From literature (PEDOTPSS data from F. Greco et al. Soft Matter, 7: 10642, 2011) or technical datasheets
[b] Experimentally measured/verified
[c] Typical value for rubber-like (incompressible) materials is 0.5, typical value for rigid polymers is 0.3
[d] Nominal value for screen printing from silver paste material datasheet
[e] Calculated

**Stand-alone devices for Tattoo Electrode interface –HW details**

Stand-alone device is built in a modular way to easily adapt to different types of measurement, having a BT microcontroller board based on AZ-BLE PSoC module (by Cypress) for data collection/elaboration and transmission connected to specific sensor boards for analog front-end for each different application. Two different devices have been designed, one specific EMG measurements and another one specific for ECG measurement. Both devices have the same BT microcontroller HW module (but with different control firmware). Both device have the same housing and assembling procedures, and can be connected with a PC via Bluetooth through a suitable receiving dongle, where transmitted data is visualised in real time by means of a custom software interface. All the material is made available as open source at the following repository:

https://doi.org/10.5281/zenodo.4382056

Three electronic board have been developed for assembling the two devices: one Bluetooth Control Board (common to the two devices, see details in Figure SI5), one EMG Sensor Board (for EMG device, see details in Figure SI6) and one ECG Sensor Board (for ECG device, see details in Figure SI7). The schematics and layouts files, developed with Eagle 6.3, are freely available at the repository (folder /HW Design), with complete bill of components.

The casing, common to the two device, has been 3D printed by stereo-lithography. STL files of device's external case (composed by two pieces, see Figure SI4) can be found at the repository (folder /HW Design) with additional details for assembling.

Once fabricate the printed circuit boards and the external case, the assembly was performed as follow:
- four magnets are inserted and glued in the holes if the bottom part of the case;
- two SMD dual-in-line pin connectors (slightly trimmed to fit in case) are glued on the bottom case;
- with conductive ink/paste are connected the pins with the magnets, to guarantee the electric connection mediated by magnetic interlocking;
- the sensor board (EMG or ECG) is inserted in the pins, and soldered to them
- the control board is staked on the Sensor board, including in the middle a small 50 mAh LiPo battery
- the battery is soldered to the control board
- the top part of the case is mounted and the device is ready for operations.

Each device is programmed with a specific firmware. Also the USB-BT dongle for PC Bluetooth connection (CY5677 CySmart BLE USB Dongle, by Cypress) need a specific firmware reprogramming. Devices' firmware and dongle's firmware are all developed with PSoC Creator 4.2 (Cypress), and are freely available at the repository.

Once programmed the devices can be connected with the PC for data transmission and acquisition. The graphic user interface software, developed in Visual Basic .NET 2017 (Microsoft), is freely available at the repository.

Finally, DXF files for ECG and EMG tattoo printing/cutting can be found at the repository. Assembly of the tattoo electrodes can be done by following the process reported in Figure SI2.

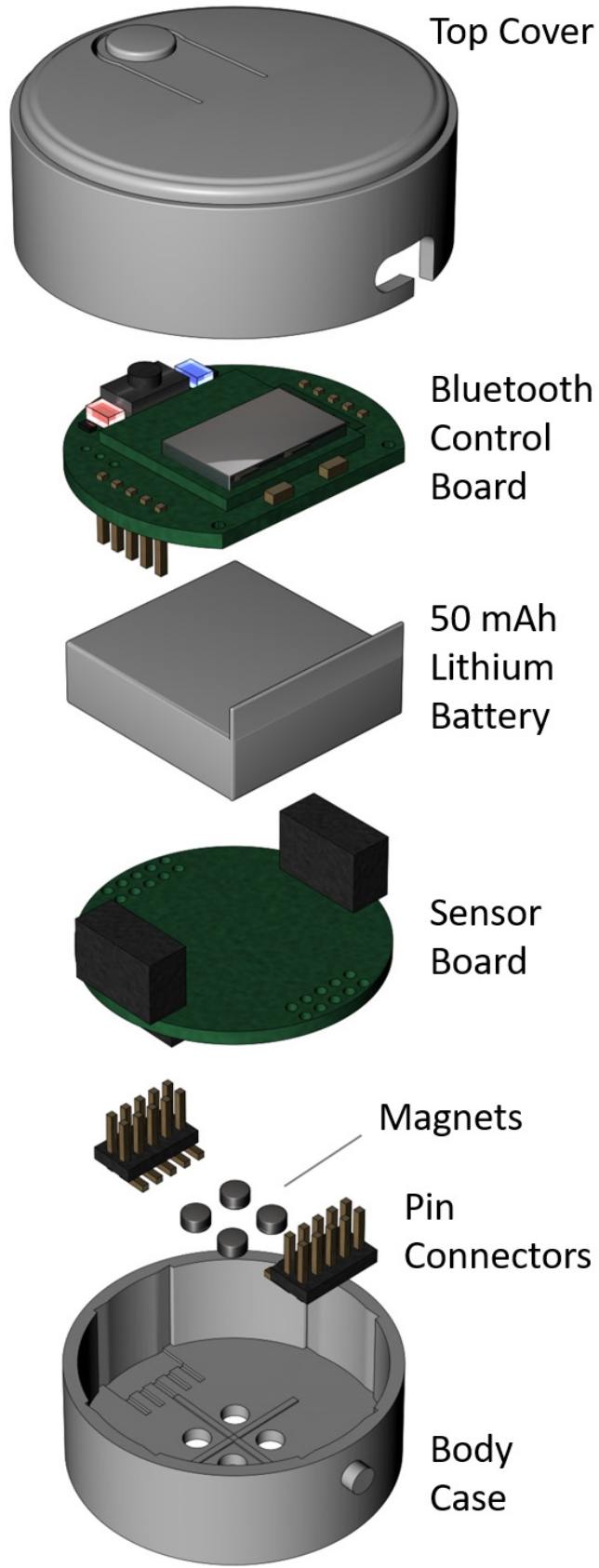
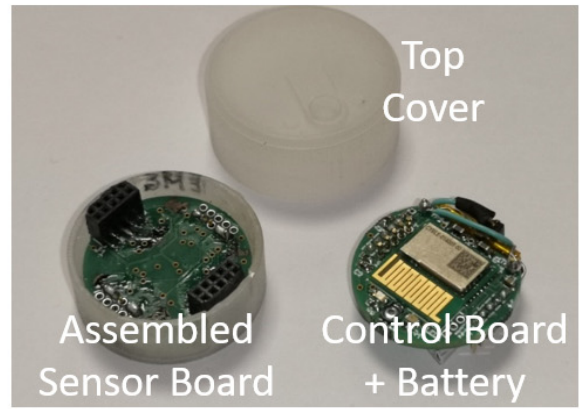
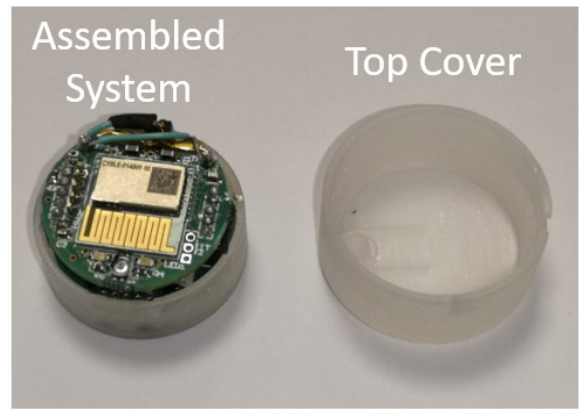
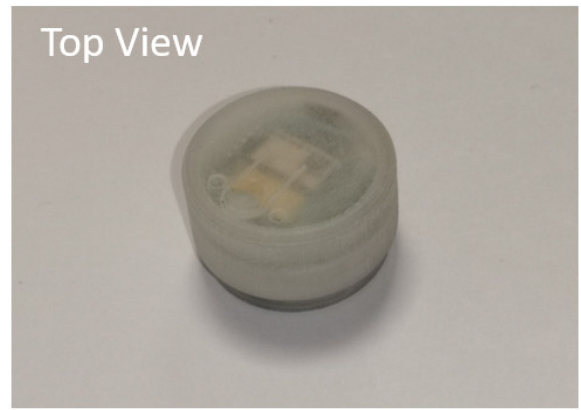
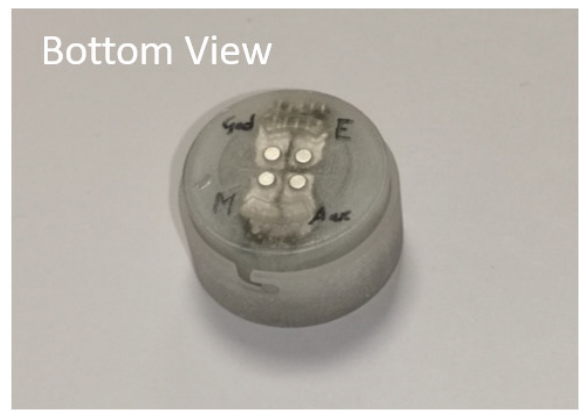

**Figure SI.4.** EMG /ECG BT device overview and pictures at different stage of assembly

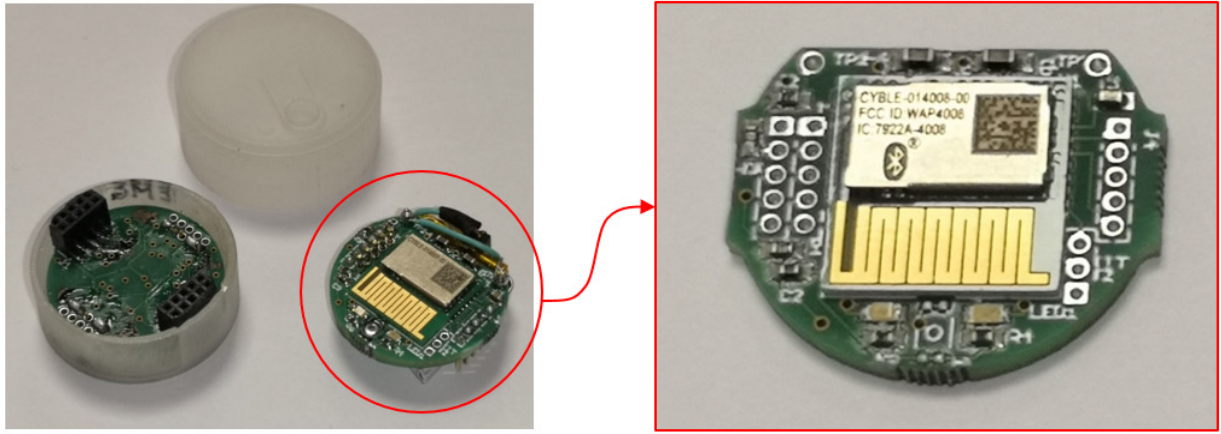

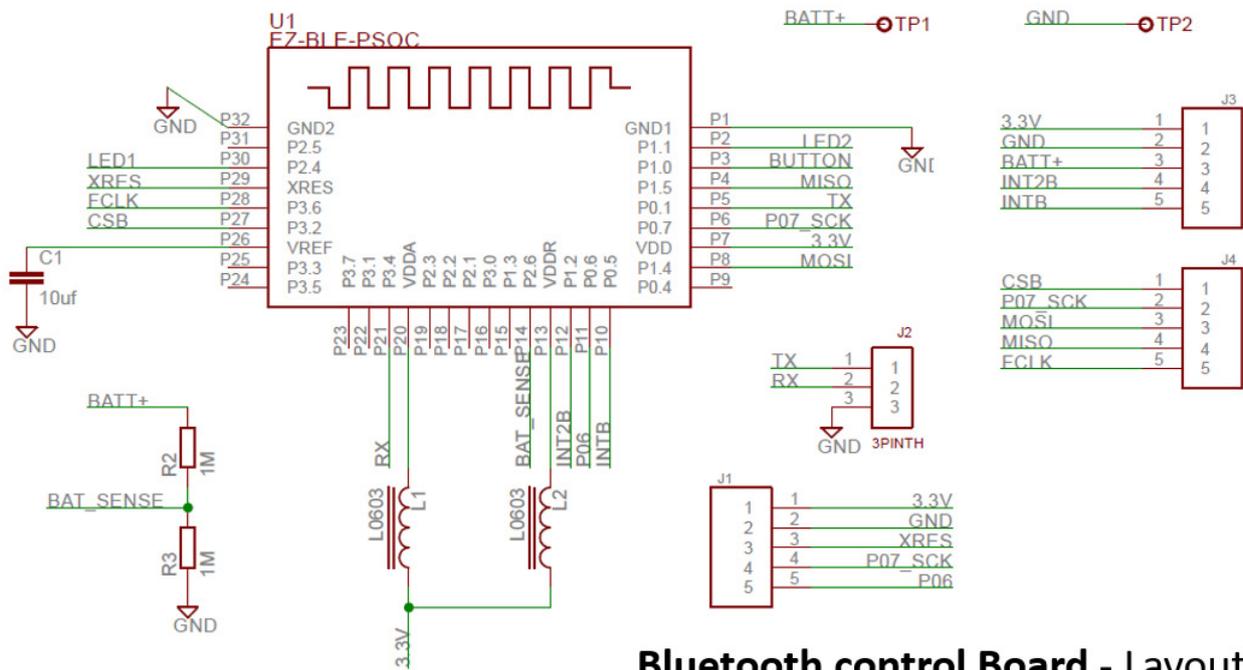

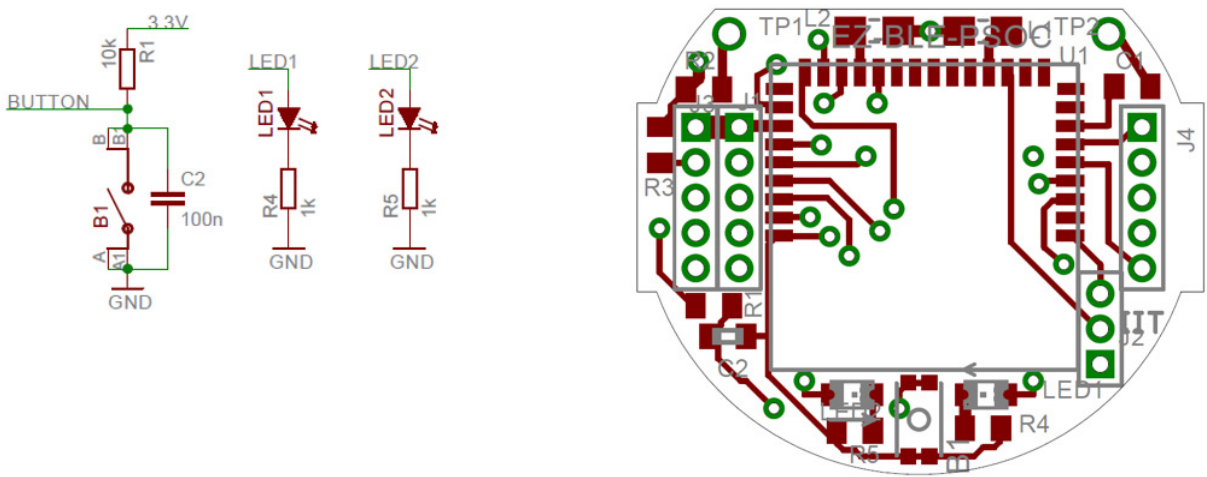

**Figure SI.5.** Bluetooth control board schematics and layout

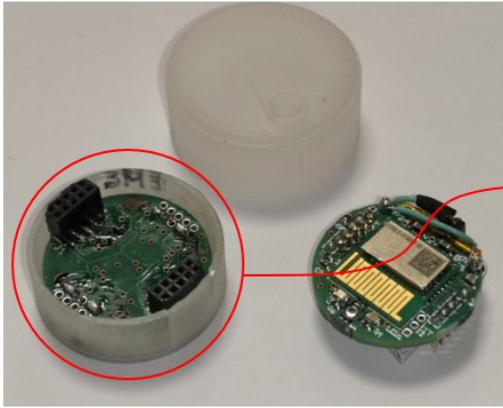
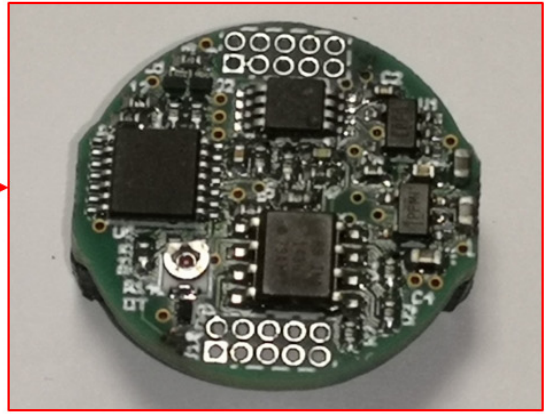

## EMG Sensor Board - Schematics

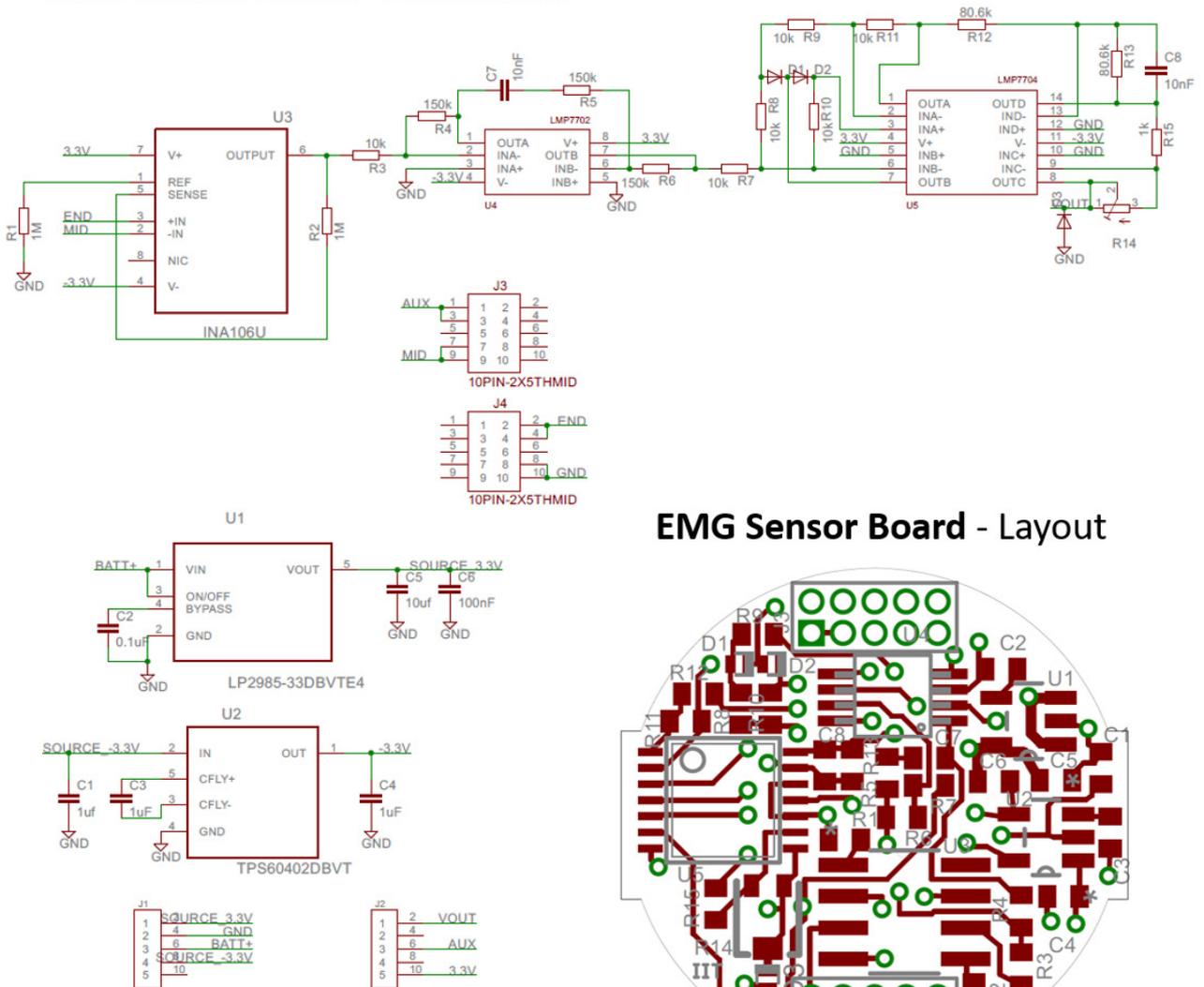

## EMG Sensor Board - Layout

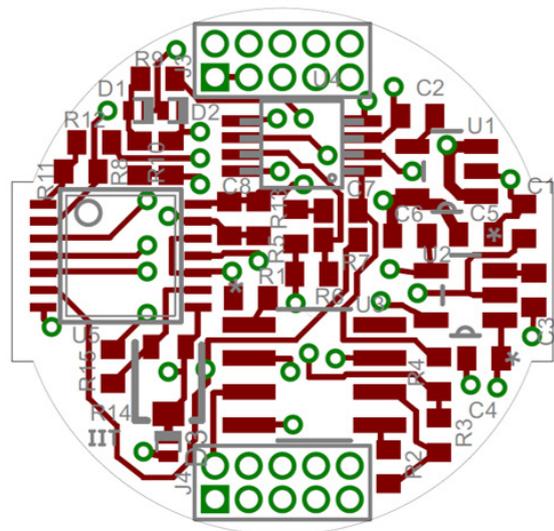

**Figure SI.6.** EMG sensor board schematics and layout - Signal conditioning schematic was partially based on the schematics available in https://www.instructables.com/id/Muscle-EMG-Sensor-for-a-Microcontroller/

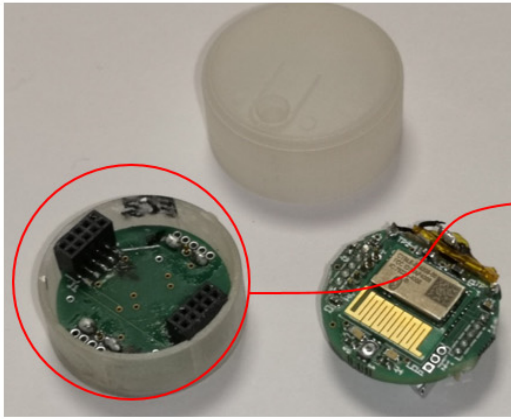
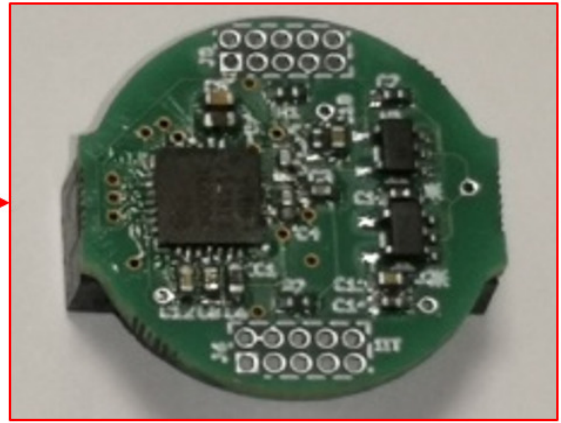
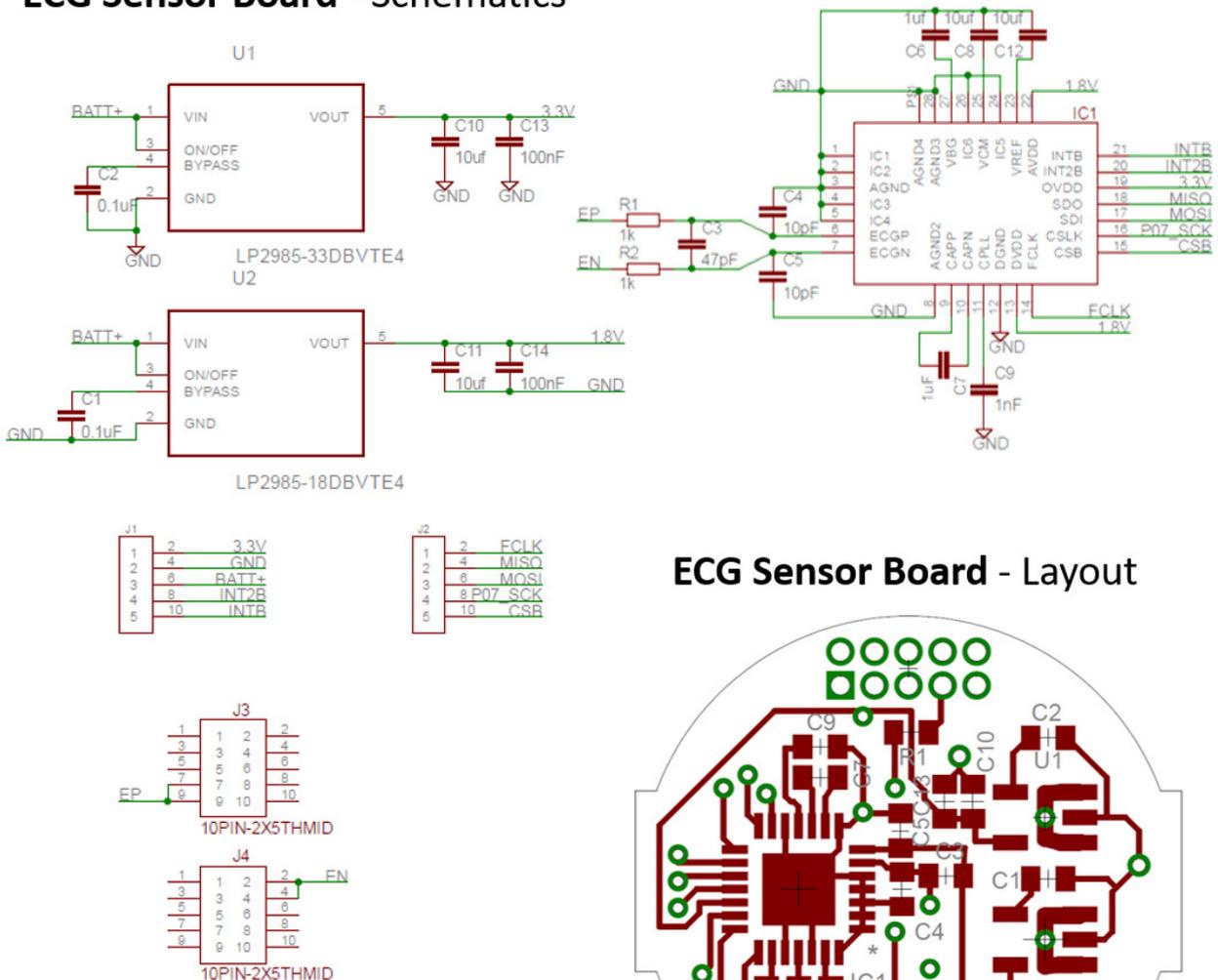
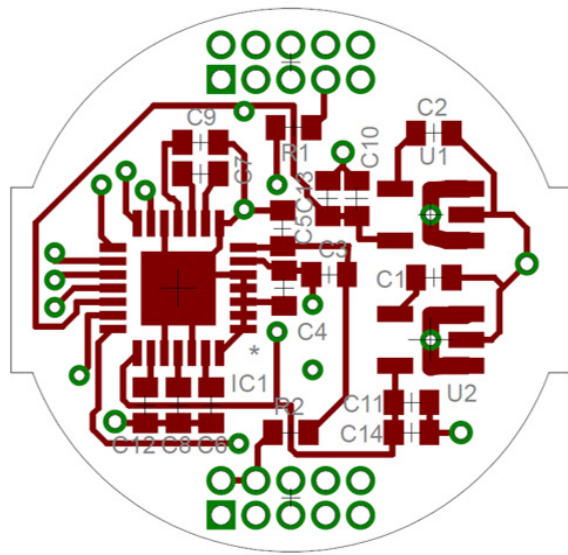

**Figure SI.7.** ECG sensor board schematics and layout – ECG signal conditioning is based on MAX30003 analog frontend

# ECG Comparison - Details

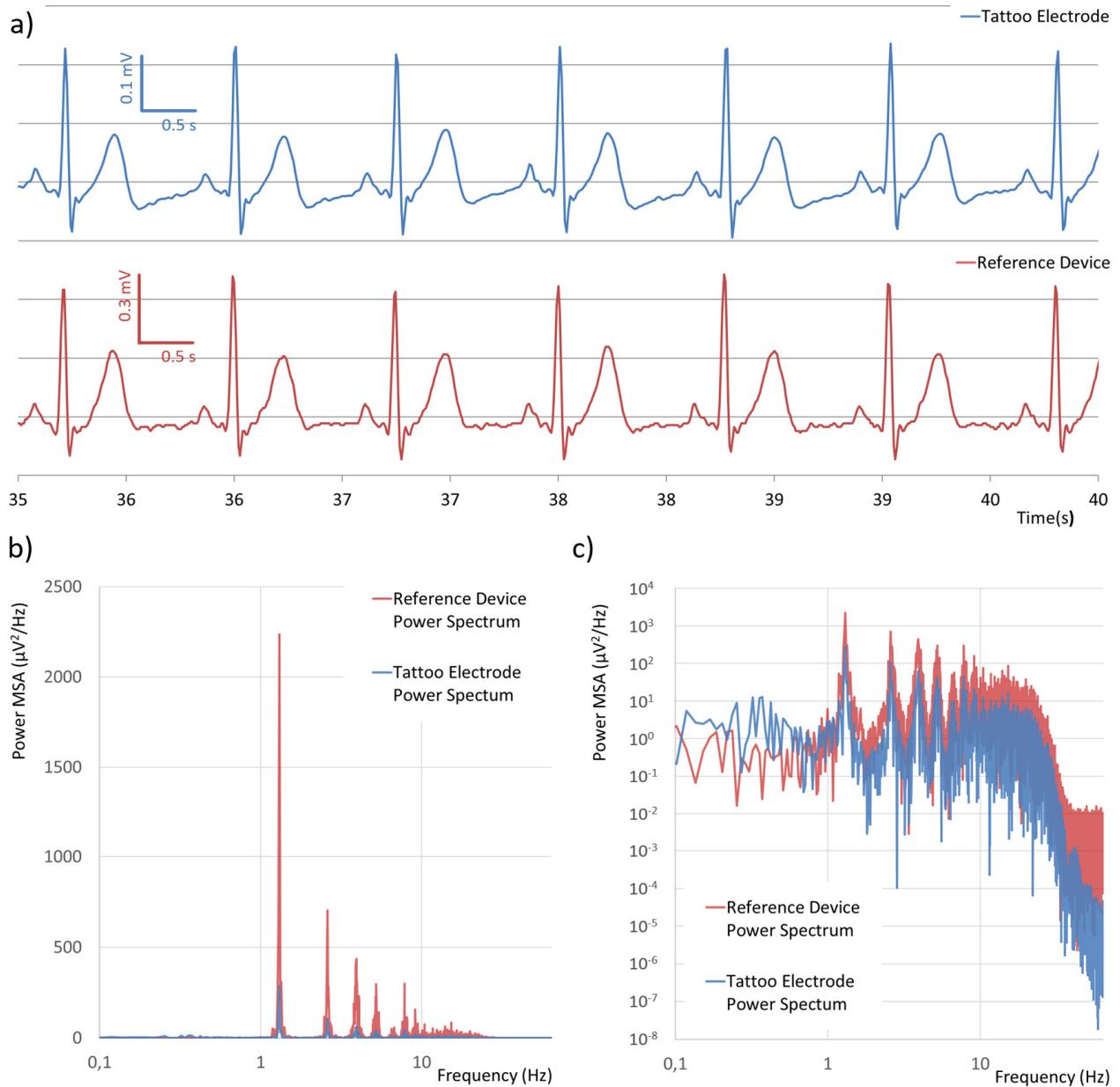

**Figure SI8.** a) Detail of ECG acquisition from tattoo electrodes by BLE device and reference device with standard electrodes; (b) Power spectrum at mean squared amplitude(MSA) of the ECG signal calculated on 60 seconds sample, both for tattoo electrode + BLE device and reference device; (c) same spectrum with different logarithmic scaling, for clarity of comparison.

**Supporting Information Video**

SI Video available for download here:
https://gitlab.iit.it/Virgilio.Mattoli/tattoo-electrode-device/-/tree/master/SI%20Videos

List of videos:

SV#1 Video of tattoo stretching test while shaking

SV#2 Tattoo electrodes release on skin with sponge and EMG signal acquisition

SV#3 EMG demo with shaking

SV#4 EMG RC car control demo

SV#5 ECG comparative demo